\documentclass[prl,amsmath,amssymb, twocolumn,superscriptaddress, 11pt,tightenlines,aps]{revtex4-1}
\usepackage{graphicx}
\usepackage{dcolumn}   
\usepackage{bm}       
\usepackage{bbm}
\usepackage{dsfont}
\usepackage{mathtools}
\usepackage{amsthm, thmtools}
\usepackage{amsmath}
\usepackage{empheq}
\usepackage{enumerate}

\newcommand{\microspace}{\mspace{0.5mu}}

\def\<{\langle}
\def\>{\rangle}
\def \lket {\left|}
\def \rket {\right\rangle}
\def \lbra {\left\langle}
\def \rbra {\right|}
\newcommand{\ket}[1]{\lket\microspace #1 \microspace\rket}

\newcommand{\bra}[1]{\lbra\microspace #1 \microspace\rbra}

\newcommand{\braket}[2]{\langle #1 \microspace | \microspace#2 \rangle}

\newcommand{\tr}[1]{\operatorname{\textnormal{Tr}}\left[ {#1} \right]} 
\def \sn {S_{\hat{n}}}
\def \heff {H_{\mathrm{eff}}}

\DeclareMathOperator{\sinc}{sinc}

\usepackage[usenames,dvipsnames]{xcolor}

\usepackage{bm}
\let\vec\bm 

\usepackage[colorlinks=true,citecolor=Cerulean,linkcolor=RubineRed,urlcolor=Cerulean]{hyperref}
\usepackage[capitalize]{cleveref}

\begin{document}
\title{Butterfly Echo Protocol for Axis-Agnostic Heisenberg-Limited Metrology\\
}
\author{Jacob Bringewatt}
\thanks{These authors contributed equally to this work.}
\affiliation{Volgenau Department of Physics, United States Naval Academy, Annapolis, MD 21402, USA}
\affiliation{Department of Physics, Harvard University, Cambridge, MA 02138, USA}
\author{Leon Zaporski}
\thanks{These authors contributed equally to this work.}
\affiliation{MIT-Harvard Center for Ultracold Atoms and Research Laboratory of Electronics, Massachusetts Institute of Technology, Cambridge, Massachusetts 02139, USA}
\author{Matthew Radzihovsky}
\affiliation{MIT-Harvard Center for Ultracold Atoms and Research Laboratory of Electronics, Massachusetts Institute of Technology, Cambridge, Massachusetts 02139, USA}
\author{Jasmine Albert}
\affiliation{Department of Physics, The College of William \& Mary, Williamsburg, Virginia 23185}
\author{Alexey V. Gorshkov}
\affiliation{Joint Center for Quantum Information and Computer Science, NIST/University of Maryland, College Park, Maryland 20742, USA}
\affiliation{Joint Quantum Institute, NIST/University of Maryland, College Park, Maryland 20742, USA}
\author{Vladan Vuleti\'c}
\affiliation{MIT-Harvard Center for Ultracold Atoms and Research Laboratory of Electronics, Massachusetts Institute of Technology, Cambridge, Massachusetts 02139, USA}
\author{Gregory Bentsen}
\affiliation{Department of Physics, The College of William \& Mary, Williamsburg, Virginia 23185}
\date{\today}
\begin{abstract}
The extreme sensitivity of chaotic systems to external perturbations makes them natural candidates for sensing applications. We propose a single-shot echo-based protocol for estimating small rotations about an unknown axis that leverages random symmetric probe states prepared via chaotic dynamics. In contrast to previous protocols for this axis-agnostic rotation sensing problem that depend on difficult-to-prepare anticoherent states, the random probe states used in our protocol can be prepared via constant-depth chaotic circuits composed of random one-axis twisting pulses. Further, the signal of interest can be extracted simply by measuring the total spin polarization. We demonstrate analytically that our protocol achieves Heisenberg scaling relative to an arbitrary rotation axis that need not be a priori known. We also investigate the effects of collective and single-particle dephasing in our protocol using analytical and numerical tools. While the requirements on dephasing rates to maintain Heisenberg sensitivity are strict, they are achievable in near-term experiments, for instance, in magnetometric rotosensing with high-spin lanthanide atoms such as $^{164} \text{Dy}$.
\end{abstract}
\maketitle

Quantum metrology harnesses many-body entanglement to enable precision measurements beyond the capabilities of uncorrelated probes~\cite{toth2014quantum}. Whereas the measurement precision of $N$ unentangled sensors is limited by the standard quantum limit (SQL) with sensitivity scaling as $\sim 1/ \sqrt{N}$, entangled probe states can surpass this limit, allowing for Heisenberg-scaling sensitivity $\sim 1/N$. This theoretical limit is difficult to achieve in practice, however. Optimal probe states, such as the Greenberger–Horne–Zeilinger (GHZ) state, are extremely sensitive to external perturbations and particle loss, motivating alternative schemes that sacrifice some amount of optimality---either in sensitivity or in signal bandwidth---to gain some degree of robustness  (although see Refs.~\cite{kielinski2024ghz,PhysRevLett.133.080801}). Examples include squeezed states~\cite{kitagawa1993squeezed,PhysRevA.50.67,giovannetti2004quantum,maccone2020squeezing}, which are less vulnerable to particle loss than GHZ states, and echo-based protocols~\cite{davis2016approaching,macri2016loschmidt,linnemann2016quantum,hosten2016quantum,nolan2017optimal,colombo2022time,kppenhofer2023squeezed,chen2024qubit}, which have intrinsic robustness to readout noise.

On the other hand, high sensitivity to external fields of interest lends itself naturally to sensing applications. In particular, chaotic systems are extremely sensitive to external perturbations due to the `butterfly effect' where small perturbations---such as a butterfly flapping its wings---are rapidly amplified into large changes in system behavior. In this Letter, we leverage this high sensitivity to develop a single-shot echo sensing protocol based on quantum information scrambling for axis-agnostic rotation sensing. The term axis-agnostic refers to our protocol's high sensitivity to the magnitude $\theta$ of rotation without requiring knowledge of the rotation axis $\hat{n}$. This feature removes the need for alignment of the probe state to a particular axis, and can potentially be leveraged to cryptographically silo rotation-axis information.

Random or chaotic dynamics have been previously considered as a metrological resource~\cite{Fiderer2018,kobrin2024universal}, and echo-based protocols have also seen significant recent interest ~\cite{davis2016approaching,macri2016loschmidt,linnemann2016quantum,hosten2016quantum,nolan2017optimal,colombo2022time,kppenhofer2023squeezed,chen2024qubit}. These two ideas have been previously combined to yield Heisenberg-limited sensing of rotations about a known axis in Ref.~\cite{kobrin2024universal}. Our work demonstrates how chaos and echo-based protocols also naturally enable optimal sensitivity for signals encoded with respect to an a priori unknown rotation axis $\hat{n}$.
These axis-agnostic rotation sensors find diverse applications, ranging from entanglement-enhanced gyroscopes for inertial navigation~\cite{goldberg2021rotation} to tests of general relativity~\cite{cerdonio1988dragging} to reference frame alignment~\cite{kolenderski2008optimal}.

\begin{figure}
\includegraphics[width=0.49\textwidth]{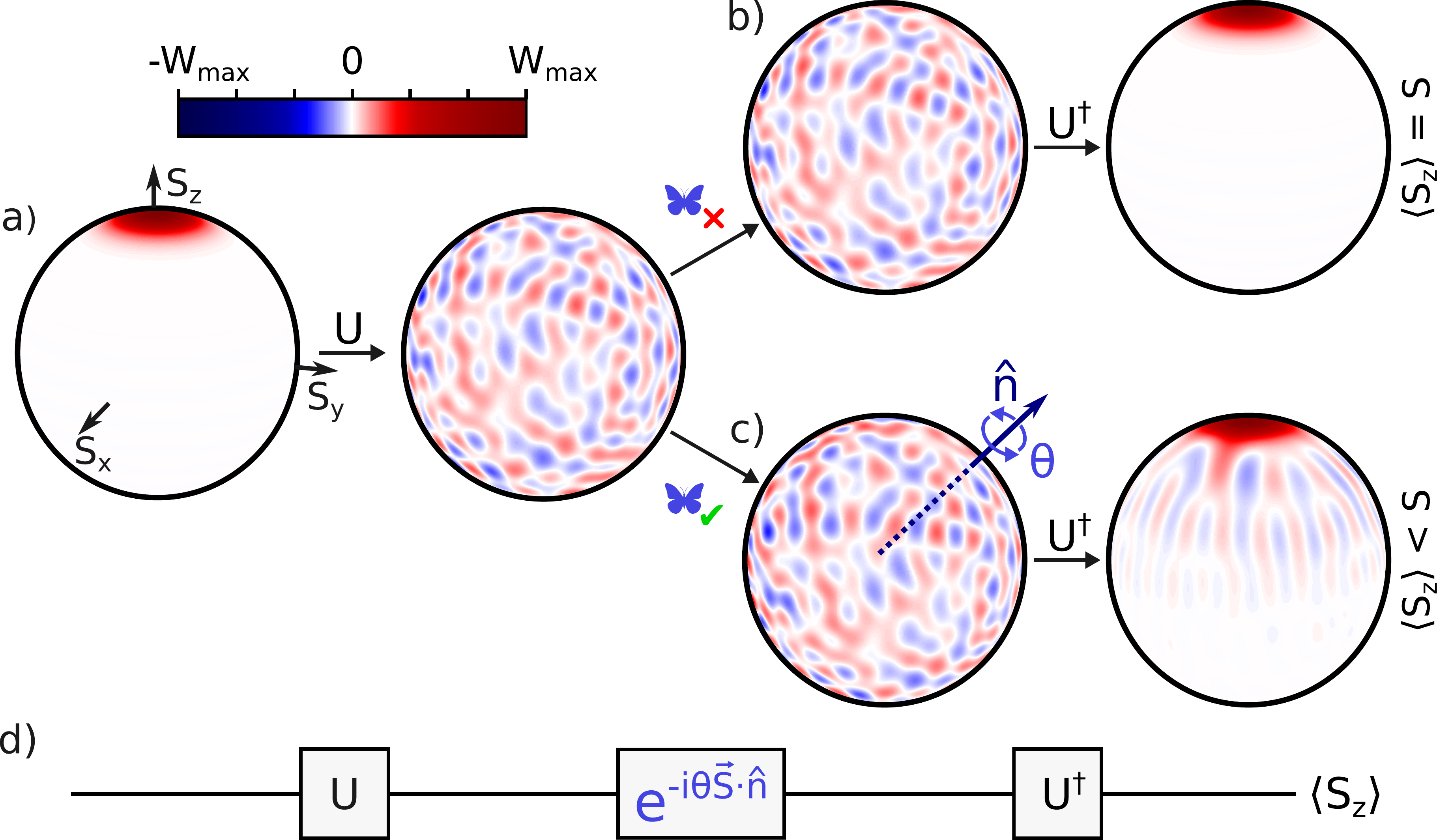}
\caption{\textbf{The butterfly echo protocol} extracts a rotation angle $\theta$ with  Heisenberg-scaling sensitivity without revealing information about the arbitrary, unknown rotation axis $\hat{n}$. An initially spin-polarized state (a) is scrambled by chaotic dynamics $U$ into a random probe state. Absent a rotation (b) the probe state returns to its original spin-polarized state under time reversal $U^{\dagger}$. By contrast, a nonzero rotation (c) leads to imperfect time-reversal and a commensurate reduction in spin polarization $\langle S_z \rangle < S$. Bloch spheres show representative Wigner functions for the quantum state at each step, with colors denoting the quasiprobability. An equivalent circuit diagram for the protocol is shown at bottom (d).}\label{fig:summary}
\end{figure}

Our work complements prior work on quantum rotosensors for this axis-agnostic sensing problem~\cite{chryssomalakos2017optimal,mo2019quantum}. For small rotation angles, the optimal probe states for rotosensing~\cite{martin2020optimal,chryssomalakos2021symmetric} are the \emph{anticoherent states}~\cite{kolenderski2008optimal,bouchard2017quantum,goldberg2021rotation,serrano2025quantum,goldberg2018quantum,zimba2006anticoherent}. However, preparing and utilizing anticoherent states for large $N$ is a significant challenge~\cite{denis2026coherent}. Here we propose an alternative \emph{butterfly echo protocol} (\cref{fig:summary}) that leverages random probe states generated by chaotic quantum dynamics.

While a straightforward application of random matrix theory demonstrates that such scrambled probe states provide Heisenberg-scaling sensitivity~\cite{oszmaniec2016random,shi2025quantum}, the problem of efficiently preparing and utilizing these states has not yet been addressed. We tackle both of these issues here: we first introduce an echo protocol that leverages quantum chaos to amplify a small rotation $\theta$ into a sharp reduction in the spin polarization $\langle S_z \rangle$; we then introduce random one-axis-twisting (OAT) circuits to prepare scrambled probe states in constant time. In particular, random OAT circuits can prepare scrambled states $\sim 1/\sqrt{N}$ faster than a GHZ state.

\textbf{\textit{Butterfly Echo Protocol.---}} Consider a system of $N$ qubit sensors prepared in a pure probe state $\ket{\psi}$, into which a small, unknown rotation angle $\theta$ is encoded via the unitary $R_{\hat{n}}(\theta) = \exp(-i{S}_{\hat{n}}\theta)$ specified by an unknown rotation axis $\hat{n}$. Here ${S}_{\hat{n}}:=\vec{S}\cdot\hat{n}$ and $\vec{S}=({S}_x,{S}_y,{S}_z) := \sum_i \vec{\sigma}_i / 2$ are the collective $\mathrm{SU}(2)$ angular momentum operators for the $N$ qubit sensors described by Pauli matrices $\vec{\sigma}_i$ with $i = 1,2,\ldots,N$.
To estimate the rotation angle $\theta$ we employ scrambled probe states $\ket{\psi} = \sum_m c_m \ket{S,m}$ that are random coherent superpositions of Dicke states $\ket{S,m}$ living in the permutation-symmetric subspace $\mathcal{S}_N$, where the $c_m$ are randomly-chosen complex coefficients, $-S \leq m \leq S$, and $S = N/2$ is the total spin.

Intuitively, these scrambled probe states are highly sensitive to arbitrary rotation axes $\hat{n}$ due to the isotropic distribution of features in the Wigner quasiprobability function (\cref{fig:summary}(a)). These scrambled probe states can thus be viewed as randomized versions of the anticoherent states used in prior rotosensing work ~\cite{goldberg2018quantum,chryssomalakos2017optimal,martin2020optimal}. In contrast to anticoherence---a notion of isotropy of a single state---the scrambled probe states considered here provide a notion of isotropy of an ensemble of states~\cite{ambainis2007quantum,gross2007evenly,roberts2017chaos}. See \cref{app:anticoherence-and-state-designs} of the supplemental material (SM) for details.

To utilize these scrambled probe states, we propose a three-step echo protocol (\cref{fig:summary}):

\textit{Step 1 (Preparation):} Starting from a completely polarized Dicke state $\ket{S,S}$, prepare the scrambled probe state $\ket{\psi} = U\ket{S,S}$ where 
$U$ is a highly chaotic random unitary acting on the symmetric subspace $\mathcal{S}_N$.

\textit{Step 2 (Encoding):} Apply the rotation $R_{\hat{n}}(\theta)$ for unknown $\hat{n}$ yielding the state $R_{\hat{n}}(\theta) \ket{\psi}$. This rotation may be regarded as a ``butterfly'' whose presence is amplified by chaotic dynamics into a large change in subsequent system behavior.

\textit{Step 3 (Measurement):} Unscramble the state via time-reversal yielding $U^{\dagger} R_{\hat{n}}(\theta) U \ket{S,S}$ and measure the spin polarization $\langle {S}_z \rangle$.

\begin{figure}
\includegraphics[width=0.49\textwidth]{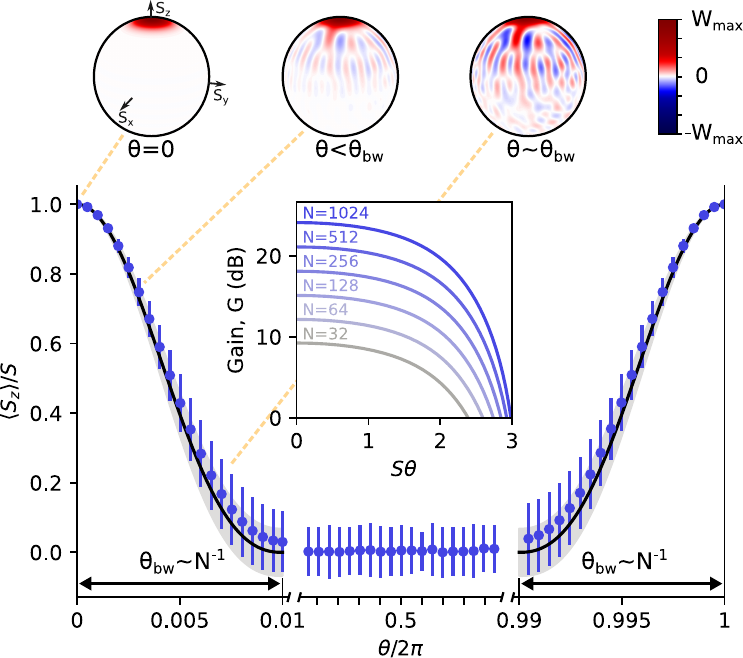}
\caption{\textbf{Spin polarization signal} $\langle S_z \rangle / S$ for $N = 100$ spins and 8 random OAT steps, each with twisting strength $\sim 1/\sqrt{N}$. Numerical simulations, averaged over 100 circuit realizations with a fixed rotation axis $\hat{n}=\hat{y}$, and analytic calculations (solid black) show a sharply decaying spin polarization signal as a function of $\theta$, with a useful metrological bandwidth scaling like $\theta_{\mathrm{bw}} \sim N^{-1}$. Shot-to-shot FWHM fluctuations in the signal (numerics in blue, analytics in grey) are subleading in the large-$N$ limit. Inset: Metrological gain for $N=32,64,\ldots,1024$ as a function of rotation angle within the bandwidth. \label{fig:fig2}}
\end{figure}

For vanishing rotation angle $\theta = 0$ (no butterfly), the forward and backward time evolutions cancel, and we are left with the original Dicke state $\ket{S,S}$ with maximal spin polarization $\langle S_z \rangle = S$. A non-vanishing rotation angle $\theta > 0$ (butterfly) leads to imperfect time-reversal and a commensurate reduction in the spin polarization $\langle S_z \rangle < S$, which serves as our metrological signal for the rotation angle $\theta$ as shown in Fig. \ref{fig:fig2}. Intuitively, our protocol is Heisenberg-limited due to the steep reduction in spin polarization, whose slope as a function of $\theta$ scales like $\sim N$ within the bandwidth $\theta_{\mathrm{bw}} \sim 1/N$.

\textbf{\textit{Axis-Agnostic Sensing at the Heisenberg Limit.---}}To analyze the butterfly echo in detail, we first investigate the problem of axis-agnostic sensing from an information-theoretic perspective. For a given rotation axis $\hat{n}$, the precision $\Delta \theta$ for estimating a rotation angle $\theta$ using a single shot is bounded by the single-parameter quantum Cram\'er-Rao bound~\cite{helstrom1976quantum,liu2020quantum}
\begin{equation}\label{eq:qcrb}
\Delta\theta\geq \frac{1}{\sqrt{\mathcal{F}_{\hat{n}}(\theta)}},
\end{equation}
$\mathcal{F}_{\hat{n}}(\theta):=4 \mathrm{Var}(\sn)$ is the Quantum Fisher Information (QFI). The QFI is upper bounded by $\mathcal{F}_{\hat{n}}(\theta)\leq N^2$ yielding an optimal single-shot sensitivity $\Delta \theta \geq 1/N$; protocols achieving this scaling with $N$ are called Heisenberg-limited. As knowing $\hat{n}$ can only improve estimation precision, any protocol for an \emph{unknown} rotation axis that manages to obtain this scaling must be optimal, up to constant factors.

A straightforward exercise demonstrates that Heisenberg scaling sensitivity is indeed obtainable in the single-shot limit even when the axis $\hat{n}$ is unknown. In particular, the encoding procedure $\rho\mapsto\rho_\theta$ for an unknown rotation axis $\hat{n}$ can be written as a mixed-unitary quantum channel
\begin{equation}\label{eq:mixed-unitary-channel}
\rho_\theta=\mathcal{M}_\theta[\rho]:=\frac{1}{4 \pi}\int d^2 \hat{n} \ R_{\hat{n}}(\theta)\rho R_{\hat{n}}^\dagger(\theta)
\end{equation}
where the integral over $\hat{n}$ expresses our ignorance of the rotation axis.
As detailed in \cref{app:channel-derivation} of the SM, for small $\theta$ this channel can be expressed as a depolarizing channel~\cite{rivas2012su2} for which we can compute the QFI, yielding $\mathcal{F}(\theta) = 4/3 \sum_j \mathrm{Var}(S_j)$ at leading order in $\theta$. Both anticoherent states and scrambled probe states maximize the variances $\mathrm{Var}(S_j) \sim S^2$ along all three spin axes $j=x,y,z$, yielding Heisenberg scaling for arbitrary unknown rotation axes $\hat{n}$.

With these general considerations in mind, we demonstrate that scrambled probe states and the butterfly echo achieve sensitivities differing from the Heisenberg limit by only constant factors. For scrambled probe states $\ket{\psi} = U\ket{S,S}$ generated by chaotic random dynamics $U$, we use random matrix theory to compute the mean QFI $\overline{\mathcal{F}_{\hat{n}}(\theta)} = N(N+1) / 3$ that is independent of the rotation axis $\hat{n}$ (see SM). This yields a single-shot sensitivity $\Delta \theta \gtrsim \sqrt{3} / N$. The overbar indicates an ensemble average over random unitary operators $U$ drawn from a random matrix ensemble acting on the symmetric subspace $\mathcal{S}_N$. While this result quantifies the typical sensitivity of scrambled probe states, one must also be concerned with fluctuations around the mean. Using similar random matrix theory tools, the variance in the QFI due to different choices of unitary operators $U$ is
\begin{equation}
\mathrm{Var}_U\left[\mathcal{F}_{\hat{n}}(\theta) \right] = \frac{4}{45} N^3 + \mathcal{O}(N^2)
\end{equation}
at large $N$, giving a standard deviation $\sim N^{3/2}$ that is subleading relative to the mean QFI.

The butterfly echo protocol achieves the same scaling up to a constant multiplicative factor. Using random matrix theory we find that the angular sensitivity of the echo protocol is
\begin{equation}\label{eq:sensitivity}
\frac{1}{\Delta{\theta}} := \left. \frac{\partial \left( S-\overline{\langle S_z \rangle} \right) / \partial \theta}{\overline{\Delta S_z}}\right|_{\theta = 0} = \frac{N}{2}+\mathcal{O}(1),
\end{equation}
where the $\mathcal{O}(1)$ terms are subleading for large $N$ \cite{collins2022weingarten}.  \Cref{eq:sensitivity} differs from optimality by only a factor of $\sqrt{3} / 2$ ~\cite{paris2009quantum}. 
 These analytic results allow us to quantify the metrological gain $G:=10\log_{10}(1/(N\Delta\theta^2))$ (inset of Figure \ref{fig:fig2}), which shows Heisenberg scaling for angles $\theta < \theta_{\mathrm{bw}}\sim N^{-1}$ inside the bandwidth.

\begin{figure}
\includegraphics[width=0.49\textwidth]{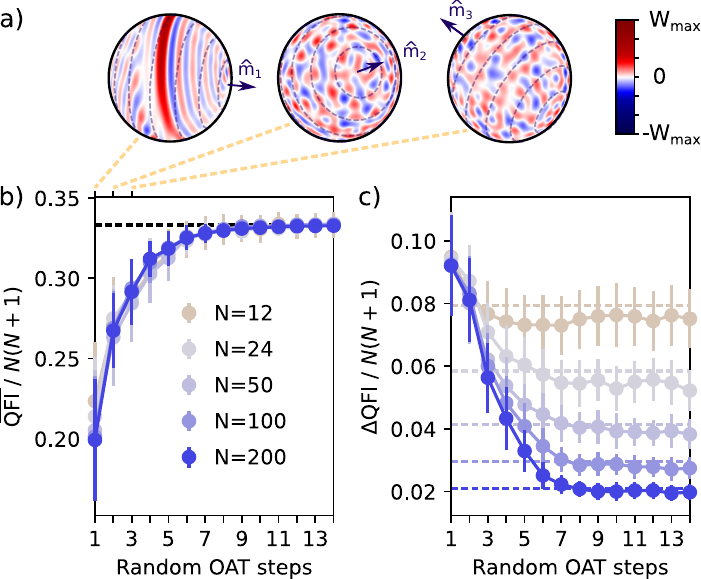}
\caption{\label{Fig1} \textbf{Probe state preparation} requires only a handful of random one-axis twisting (OAT) pulses. Successive applications of random OAT pulses yield approximately random probe states (a) with a mean QFI (b, dots) that rapidly approaches the ideal random-matrix value $\overline{\mathrm{QFI}} = N(N+1)/3$ (dashed) on a timescale that is independent of the system size $N$. 
Fluctuations in the QFI (c, dots) also converge to their random-matrix values (dashed) on the same timescale. Numerical data is averaged over 100 random OAT circuits (300 for $N=12$) and $10^3$ randomly-chosen rotation axes $\hat{n}$. Error bars are shrunk by a factor of 3 for clarity. }
\label{fig:fig3}
\end{figure}

\textbf{\textit{Probe State Preparation.---}}While ideal scrambled probe states are prepared using random unitary operators $U$ acting on the symmetric subspace $\mathcal{S}_N$, such operators are extremely difficult to generate in realistic experiments. Here we show that low-depth chaotic random circuits suffice to generate approximate random unitary dynamics (\cref{fig:fig3}) using only rotations and one-axis twisting (OAT) operations that are native to relevant experimental platforms including cavity QED~\cite{PhysRevLett.104.073602}, trapped ions~\cite{doi:10.1126/science.aad9958}, and NV centers \cite{Wu2025}.
Inspired by the canonical kicked top model \cite{haake1987classical,yin2021quantum,Fiderer2018,wang2011chaos}, we consider an ensemble of random circuits composed of a series of one-axis twisting pulses $U_j = \exp \left( - i \chi t S^2_{\hat{m}_j} \right)$ performed along random axes $\hat{m}_j$ at each timestep $j = \{1,2,\ldots\}$. Each pulse applies a twisting strength $\chi t = \tfrac{\pi}{2\sqrt{N}}$ corresponding to the strength required to wrap the Wigner function of a coherent state around the Bloch sphere once (see Fig. \ref{fig:fig3}(a)).

These chaotic random OAT circuits rapidly scramble the spin and generate approximately random probe states with Heisenberg-scaled performance after only a handful of twisting operations (Fig. \ref{fig:fig3}). Visual inspection of the Wigner functions at each timestep $t$ (Fig. \ref{fig:fig3}(a)) already provides an indication of convergence to randomness. Direct calculation of the mean QFI---averaged over circuit realizations and rotation axes (Fig. \ref{fig:fig3}(b), dots)---reveals rapid convergence to the ideal random-matrix value $N(N+1) / 3$ (dashed). Crucially, the number of pulses required for convergence is constant in the system size $N$ (by contrast, preparing a GHZ state requires $\mathcal{O}(\sqrt{N})$ OAT steps). To verify this we have also analytically studied a related infinitesimal random OAT model in which the twisting strength $\chi dt$ per pulse is taken to zero while the total twisting strength is fixed to $\chi t = c / \sqrt{N}$ to match the numerical model (see SM, \cref{app:HeffDerivation}). The ensemble-averaged effective Hamiltonian for this analytical model features an energy gap that is constant with system size, indicating a timescale to convergence that does not depend on $N$. 

While the mean QFI provides a measure of the \emph{typical} metrological usefulness of probe states prepared by random OAT dynamics, one must also be concerned with the size of fluctuations around the mean due to different realizations of random preparation circuits and to variations in the rotation axis $\hat{n}$. Numerical simulations (Fig. \ref{fig:fig3}(c)) demonstrate rapid convergence of the standard deviation of the QFI (dots) to its random-matrix value $\Delta \mathrm{QFI} = \tfrac{2 N^{3/2}}{3\sqrt{5}}$ (dashed) on the same timescale as the mean QFI. Again, the number of timesteps required for convergence is independent of system size $N$.

\begin{figure}
\includegraphics[width=0.45\textwidth]{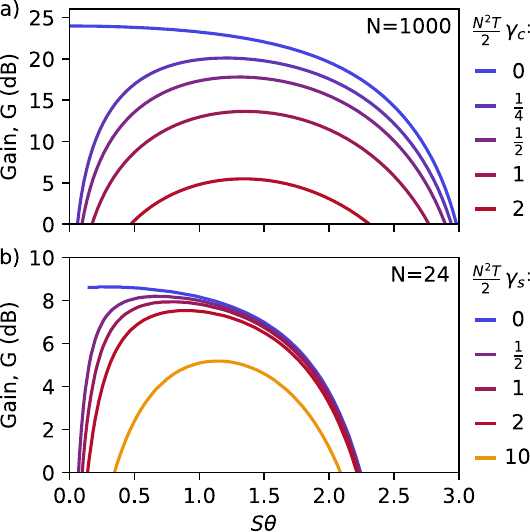}
\caption{\label{Fig4}
\textbf{Decoherence reduces gain and bandwidth} of the butterfly echo protocol. We consider increasing rates of (a) collective and (b) single-particle dephasing, expressed in terms of the system size, $N$, and the total random one-axis twisting evolution time $T{\sim} \frac{\pi}{2\chi\sqrt{N}} \times \mathcal{O}(1)$. The results in (a) are analytic for $N=10^3$ and scale universally with $N$, whereas the results in (b) are numerical for $N=24$ and $T=\frac{\pi}{2\chi \sqrt{N}}\times 8$.
}\label{fig:fig4}
\end{figure}

\textbf{\textit{Effects of Decoherence.---}}With an eye toward experimental implementation, we also study the sensitivity of the butterfly echo protocol to collective and single-particle dephasing (\cref{fig:fig4}). Analytic calculations combining random matrix theory and degenerate perturbation theory (SM, \cref{app:noisypert}) yield an analytic expression for the metrological gain in the presence of collective dephasing at a rate $\gamma_c$ (Fig. \ref{fig:fig4}(a)). Although decoherence completely destroys the metrological gain at vanishingly small rotation angles $\theta \sim 0$, we still find sensitivities surpassing the standard quantum limit for a substantial fraction of the bandwidth $\theta_{\textrm{bw}}$ up to collective dephasing rates of $\gamma_c{\sim}\chi N^{-3/2}\times \mathcal{O}(1)$. The rotation angle for which the metrological gain is maximal is located at roughly half the bandwidth, which makes intuitive sense given that the signal $\langle S_z \rangle$ has its steepest slope at around $\theta \sim \theta_{\mathrm{bw}} / 2$ as shown in \cref{fig:fig2}.

We observe qualitatively similar behavior for increasing levels of single-particle dephasing at a rate $\gamma_s$. Because such a process destroys the symmetry of the state, we numerically analyze a few-body system using the Permutation Invariant Quantum Solver from QuTiP~\cite{qutip}.  The numerical results in \cref{fig:fig4}(b) indicate that our protocol is more resilient to single-particle dephasing than collective dephasing.

The requirements on dephasing rates are strict, and they further tighten with an increase in particle number $N$. Nevertheless, the butterfly echo protocol can still lead to a practical advantage in magnetometric rotosensing with high-spin lanthanide atoms such as dysprosium-164. Tensor-light–shift–enabled OAT in dysprosium produces intrinsic spin squeezing, as well as intrinsic non-Gaussian~\cite{PhysRevLett.122.173601} and GHZ states~\cite{Chalopin2018}, demonstrating sufficient levels of coherence for our protocol to be effective.

\textbf{\textit{Outlook.---}}Although random OAT circuits are capable of generating sufficiently scrambled probe states in constant time, recent results in the study of unitary $k$-designs indicate that there may be more efficient methods to prepare the necessary probe states ~\cite{laracuente2024approximate,schuster2025random,grevink2025will}. Further, end-users nescient of the specific random unitary $U$ used to prepare the probe state have zero knowledge about the rotation axis $\hat{n}$. This ability to effectively hide the rotation axis from users could be leveraged to cryptographically silo information in blind navigation protocols -- a metrological counterpart of blind quantum computing \cite{Fitzsimons2017} -- and suggests deeper connections between metrology, scrambling, and cryptography. We leave these interesting questions for future work.

\begin{acknowledgments}
\textbf{{\textit{Acknowledgments.---}}}We thank Brian Swingle for helpful discussions early in the development of this work. We also thank Irina Novikova for helpful discussions regarding the echo protocol's robustness to decoherence and for comments on the manuscript. JB notes that the views expressed in this work are those of the author and do not reflect the official policy or position of the United States Naval Academy or any department of the United States Government. L.Z., M.R., A.V.G., and V.V.~acknowledge support from the U.S.~Department of Energy, Office of Science, National Quantum Information Science Research Centers, Quantum Systems Accelerator (award No.~DE-SCL0000121). A.V.G.~was also supported in part by ONR MURI, ARL (W911NF-24-2-0107), the DoE ASCR Quantum Testbed Pathfinder program (award No.~DE-SC0024220), NSF QLCI (award No.~OMA-2120757), NSF STAQ program, AFOSR MURI,  DARPA SAVaNT ADVENT, and NQVL:QSTD:Pilot:FTL. A.V.G.~also acknowledges support from the U.S.~Department of Energy, Office of Science, Accelerated Research in Quantum Computing, Fundamental Algorithmic Research toward Quantum Utility (FAR-Qu).

\end{acknowledgments}

\textit{\textbf{Note added:} In the final stages of preparation of this manuscript, Ref.~\cite{liu2026echoed} was posted on arXiv, presenting numerical results on a similar echo-based protocol. Their results are consistent with ours.
}

\bibliography{References}

\onecolumngrid

\newpage
\begin{center}
\textbf{Supplemental Material} 
\end{center}
\setcounter{secnumdepth}{1}
\renewcommand{\thesection}{S\arabic{section}}
\setcounter{theorem}{0}
\renewcommand{\thetheorem}{S\arabic{theorem}}
\setcounter{lemma}{0}
\renewcommand{\thelemma}{S\arabic{lemma}}
\setcounter{equation}{0}
\renewcommand{\theequation}{S\arabic{equation}}
\setcounter{table}{0}
\renewcommand{\thetable}{S\arabic{table}}
\setcounter{figure}{0}
\renewcommand{\thefigure}{S\arabic{figure}}

In this supplemental material we elaborate on the details of the calculations described in the main text. In particular, \cref{app:anticoherence-and-state-designs} we elaborate on the connection between the anticoherent states that have previous been studied in the context of the rotosensing problem and quantum state designs over the symmetric space $\mathcal{S}_N$.  In \cref{app:channel-derivation} we show how the mixed unitary channel in \cref{eq:mixed-unitary-channel} of the main text, describing rotation about an unknown axis, can be expressed as a depolarizing channel. We then compute the quantum Fisher information for this channel, indicating that Heisenberg scaling is, indeed, possible for the rotosensing problem. In \cref{app:analysis} we provide the details of the random matrix analysis that rigorously demonstrate the performance of our butterfly echo protocol when using Haar random symmetric states. In \cref{app:noisypert} we provide details for the analytic calculation of the effects of collective dephasing on the butterfly echo protocol. In \cref{app:HeffDerivation} we analyze a Brownian circuit model for generating random symmetric probe states and show that it is equivalent to the random OAT model for preparing probe states in the limit of infinitesimal twists. Examining the spectrum of the effective Hamiltonian that describes this model indicates that the timescale to reach a metrologically useful Haar-random probe state is constant in system size. Finally, in \cref{app:mmse} we compare our estimation scheme (measuring $S_z$) to the minimum mean square error (MMSE) estimator that is optimal from a Bayesian approach to analyzing the rotosensing problem.

\section{Anticoherence versus Symmetric State Designs}\label{app:anticoherence-and-state-designs}
In the section, we elaborate on the connections between anticoherent states and symmetric quantum state designs. As described in the main text, both are, in some sense, isotropic (with respect to the rotation axis $\hat{n}$). However, while anticoherence is a property of a single state, a symmetric state design is an ensemble of states.

A quantum state $\rho$ is order-$k$ anticoherent if $\langle \hat{S}_{\hat{n}}^t\rangle$ is independent of $\hat{\vec{n}}$ for all $t\leq k$~\cite{zimba2006anticoherent,martin2020optimal}. Thus, the order of anticoherence specifies the extent of isotropy of a state: a higher order anticoherent state requires looking at expectation values of higher moments of spin operators to see any anisotropy. From this definition and the expression for the quantum Fisher information for rotation about a known axis $\hat{n}$,
\begin{equation}\label{eq:qfi}
\mathcal{F}_{\hat{n}}(\theta):=4(\Delta{S}_{\hat{n}})^2=4\left[\langle {S}_{\hat{n}}^2\rangle - \langle\hat{S}_{\hat{n}}\rangle^2\right],
\end{equation}
it follows that an order-2 anticoherent state provides Heisenberg scaling $\sim S^2$ for measuring a rotation about any \emph{known} axis~\cite{goldberg2018quantum}. 

When considering rotosensing with an unknown axis, it is standard practice to examine the fidelity of the rotated state with the probe state averaged over $\hat{n}$; with this figure of merit (which implies a particular choice of $\hat{n}$-independent measurement), higher order anticoherence always improves the sensitivity of the rotosensor~\cite{chryssomalakos2017optimal,martin2020optimal}.

Recall that a symmetric quantum state $k$-design is an ensemble of states $\mathcal{E}\subseteq\mathcal{S}_N$ such that $\overline{f_k(\rho)} = \mathbb{E}_{\rho\in\mathcal{E}}\left[f_k(\rho)\right]$, for any polynomial of degree $k$, $f_k(\rho)$, where the overbar indicates the Haar average. By the isotropy of the symmetric Haar ensemble, this definition implies, that $\overline{\prod_{j=1}^t\langle{S}_{\hat{\vec{n}}}^{s_j}\rangle}$ is independent of $\hat{n}$ for all $t\leq k$ and all $s_j\in\mathbb{Z}^+$. 

Where anticoherence is a notion of isotropy of an individual state, quantified by the isotropy of spin moments, symmetric state designs provide a notion of isotropy of an ensemble of states, quantified by $k$-body correlators of arbitrary spin moments. Nonetheless, there is clearly a correspondence between the two notions of isotropy. For instance, where an order-2 anticoherent state implies an isotropic $\mathcal{F}_{\hat{n}}(\theta)$ with Heisenberg scaling, a state $2$-design implies an isotropic ensemble-averaged QFI, $\overline{\mathcal{F}_{\hat{n}}(\theta)}$, with Heisenberg scaling.  In this sense, a symmetric state $k$-design gives a notion of order-$k$ anticoherence on average, providing intuition for the success of the butterfly echo protocol

\section{Mixed Unitary Channel Form of the Unknown Axis Problem} \label{app:channel-derivation}

\subsection{From Mixed Unitary Channel to Depolarizing Channel}
In this section, we show that the mixed-unitary channel in \cref{eq:mixed-unitary-channel} describing the encoding procedure for the rotosensing problem can be expressed as a depolarizing channel. In particular, any non-adaptive scheme or when operating in the single-shot regime, a rotosensing protocol must be rotation-axis agnostic. Thus, as described in the main text we can consider the encoding procedure as a mixed-unitary quantum channel
\begin{align}\label{eq:channel}
\rho_\theta = \mathcal{M}_\theta[\rho]&:=\int d\hat{n} \, p(\hat{n}) R_{\hat{n}}(\theta)\rho R^\dagger_{\hat{n}}(\theta)\nonumber\\
&=\frac{1}{4\pi}\int d\hat{n} \,  R_{\hat{n}}(\theta)\ket{\psi}\bra{\psi} R^\dagger_{\hat{n}}(\theta)
\end{align}
where $p(\hat{n})$ is our prior distribution for the unknown rotation axis, which we take to be a uniform distribution, $\rho=\ket{\psi}\bra{\psi}$ is the probe state, and $R_{\hat{n}}(\theta)$ is a unitary rotation about the axis $\hat{n}$ by angle $\theta$.

For small $\theta$ we can write the mixed-unitary channel in \cref{eq:channel} in a somewhat simpler form,
\begin{align}\label{eq:channel-new}
\mathcal{M}_\theta[\rho] &= \rho + \mathcal{D}_\theta[\rho] + \mathcal{O}(\theta^3),
\end{align}
where $\mathcal{D}_\theta[\rho]$ is the depolarizing channel
\begin{align}\label{eq:depolarization}
\mathcal{D}_\theta[\rho]:= \frac{\theta^2}{3}\sum_{j\in\{x,y,z\}}\Big(S_j \rho S_j-\frac{1}{2}\{S_j^2,\rho\}\Big)
\end{align}

To see this, simply expand \cref{eq:channel} to second order in $\theta$:
\begin{align}
\mathcal{M}_\theta[\rho] = \rho &+ \frac{1}{4\pi}\int d\hat{n} \left[-i\hat{n}\cdot\vec{S}\rho + i \rho \hat{n}\cdot\vec{S}\right]\theta \nonumber \\
&+ \frac{1}{4\pi}\int d\hat{n} \left[(\hat{n}\cdot\vec{S})\rho(\hat{n}\cdot\vec{S}) - \frac{1}{2}\rho (\hat{n}\cdot\vec{S})^2 - \frac{1}{2}(\hat{n}\cdot\vec{S})^2\rho\right]\theta^2 + \mathcal{O}(\theta^3).
\end{align}
Using spherical harmonics it is straightforward to show that
\begin{subequations}
\begin{align}
\int d\hat{n} \, \hat{n} &= 0, \\
\int d\hat{n} \, \hat{n}_i\hat{n_j} &= \frac{4\pi}{3}\delta_{ij},
\end{align}
\end{subequations}
so the terms linear in $\theta$ vanish and we obtain \cref{eq:channel-new}.

\subsection{Quantum Fisher Information for the Depolarizing Channel}\label{app:qfi-for-depolarizing-channel}
Now that we have reduced the problem of learning a small rotation $\theta$ about an unknown axis to estimating the strength of a depolarization channnel $\mathcal{M}_\theta[\rho]:=\rho+\mathcal{D}_\theta[\rho]+\mathcal{O}(\theta^3)$, we compute the quantum Fisher information for this channel.

For small $\theta$, the quantum Fisher information can be expressed as
\begin{equation}
\theta^2\mathcal{F}(\theta) = 8\left(1-\sqrt{f(\rho,\mathcal{M}_\theta[\rho]})\right),
\end{equation}
where 
\begin{equation}
f(\rho,\sigma)=\mathrm{Tr}\left[\sqrt{\sqrt{\rho}\sigma\sqrt{\rho}}\right]^2,
\end{equation}
is the fidelity. For a pure probe state $\rho=\ket{\psi}\bra{\psi}=\rho^2$, one finds that
\begin{align}
f(\rho,\mathcal{M}_\theta[\rho])&=\mathrm{Tr}\left[\sqrt{\rho +\frac{\theta^2}{3}\left(\sum_j \langle S_j\rangle^2\rho-\langle S_j^2\rangle\rho\right)+\mathcal{O}(\theta^3)}\right]^2 \nonumber \\
&=\mathrm{Tr}\left[\rho\sqrt{1 +\frac{\theta^2}{3}\left(\sum_j \langle S_j\rangle^2-\langle S_j^2\rangle\right)+\mathcal{O}(\theta^3)}\right]^2 \nonumber\\
&=\left[1-\frac{\theta^2}{6}\sum_j\mathrm{Var}(S_j)+ \mathcal{O}(\theta^3)\right]^2.
\end{align}
Thus, 
\begin{equation}
\mathcal{F}(\theta) = \frac{4}{3} \sum_j \mathrm{Var}(S_j)+\mathcal{O}(\theta^3).
\end{equation}
Thus, for a Haar random symmetric probe state, the expectation value of the quantum Fisher information is
\begin{equation}
\overline{\mathcal{F}(\theta)} = \frac{4}{3} S(S+1)+\mathcal{O}(\theta) \sim S^2,
\end{equation}
where we use Eq.~(\ref{eq:qfivar}) from Sect.~\ref{app:analysis}.
The quantum Cram\'er-Rao bound implies that the signal to noise ratio $R$ (for a single measurement) using a Haar random symmetric state is bounded (on average) as~\cite{paris2009quantum}
\begin{equation}
R^2:=\frac{\theta^2}{\mathrm{Var}(\tilde\theta)} \leq \theta^2 \overline{\mathcal{F}(\theta)} \sim \theta^2 S^2
\end{equation}
The butterfly-echo protocol yields an average signal-to-noise ratio $R \sim S\theta$, consistent with the optimal scaling as determined via the quantum Cram\'er-Rao bound.

\section{Random Matrix Analysis for Butterfly Echo Protocol}\label{app:analysis}

In this section, we provide the details of the rigorous analysis of the butterfly echo protocol using random matrix theory. We initialize the system in the completely polarized Dicke state $\ket{S,S}$ and prepare the probe state $\ket{\psi} = U \ket{S,S}$ by applying a random $D \times D$ unitary matrix $U$ sampled from the Haar ensemble $\mathcal{H}$ \cite{collins2022weingarten}, where $D = 2S+1$ is the Hilbert space dimension. Initially we will focus on the clean (noiseless) case where $\gamma = 0$, but in later sections we will generalize to account for depolarizing noise that acts during state preparation and time-reversal.

\subsection{Expectation Values and Quantum Fisher Information: First Moment Analysis}

Before considering the echo protocol, let us first prepare the probe state $\ket{\psi}$ and study ensemble-averaged expectation values
\begin{equation}
    \overline{\bra{\psi} \mathcal{O} \ket{\psi}} = \mathbb{E}_{U \sim \mathcal{H}} \left[ \bra{S,S} U^{\dagger} \mathcal{O} U \ket{S,S} \right]
\end{equation}
where the overbar indicates an average over Haar random unitaries $U$ and $\mathcal{O}$ is an arbitrary operator acting on the symmetric Dicke subspace. Such quantities are categorized as first-moment ($k = 1$) quantities because they involve a single forward evolution $U$ and a single backward evolution $U^{\dagger}$. Using the Choi-Jamio\l kowski isomorphism (see Figure \ref{fig:Choi-Jam}) and the Weingarten calculus, we find that expectation values are given by the trace
\begin{equation}
    \overline{\bra{\psi} \mathcal{O} \ket{\psi}} = \tr{\mathcal{O}} / D
\end{equation}
where $D = 2S+1$ is the dimension of the Hilbert space. In other words, ensemble-averaged expectation values are just the expectation values for a maximally mixed state $\rho = \mathbb{I} / D$. This makes sense because the ensemble average over random unitaries $U$ generates a maximally mixed state:
\begin{equation}
    \overline{U \ket{S,S} \bra{S,S} U^{\dagger}} = \mathbb{I} / D.
    \label{eq:maxmixed}
\end{equation}
In particular, the average spin polarization vanishes in all directions $\overline{\bra{\psi} \sn \ket{\psi}} = \tr{\sn}/D = 0$ and the variance is isotropic in the spin axis $\hat{n}$:
\begin{equation}
    \overline{\bra{\psi} S_{\hat{n}}^2 \ket{\psi}} = \tr{\sn^2} / D = \frac{1}{3}S(S+1).
\end{equation}
This calculation captures the metrological usefulness of the probe state $\ket{\psi} = U \ket{S,S}$ because the variance governs the quantum Fisher information (QFI):
\begin{equation}
    \overline{\mathcal{F}_{\hat{n}}(\theta)} = 4 \overline{\mathrm{Var}[\sn]} = \frac{1}{3} N (N+1)
    \label{eq:qfivar}
\end{equation}
which differs from the Heisenberg limit by a factor of $1/3$.
yielding Heisenberg-limited scaling $\sim N^2$. Getting the correct subleading term $N/3$ above actually requires a $k = 2$ calculation, which we turn to in the next section.

Note that Eq. \eqref{eq:qfivar} only holds for pure states $\ket{\psi}$ whereas a more complicated expression applies for mixed states~\cite{liu2020quantum}. This may cause some confusion since we just mentioned the maximally mixed state in Eq. \eqref{eq:maxmixed}. However the maximally mixed state only appears after averaging over the ensemble $U \sim \mathcal{H}$; prior to this averaging every probe state $\ket{\psi}$ is pure and we may therefore use the simple pure-state expression Eq. \eqref{eq:qfivar} for the QFI.

\subsection{Echo Protocol: Second Moment Analysis}

\begin{figure}
    \centering
    \includegraphics[width=0.7\linewidth]{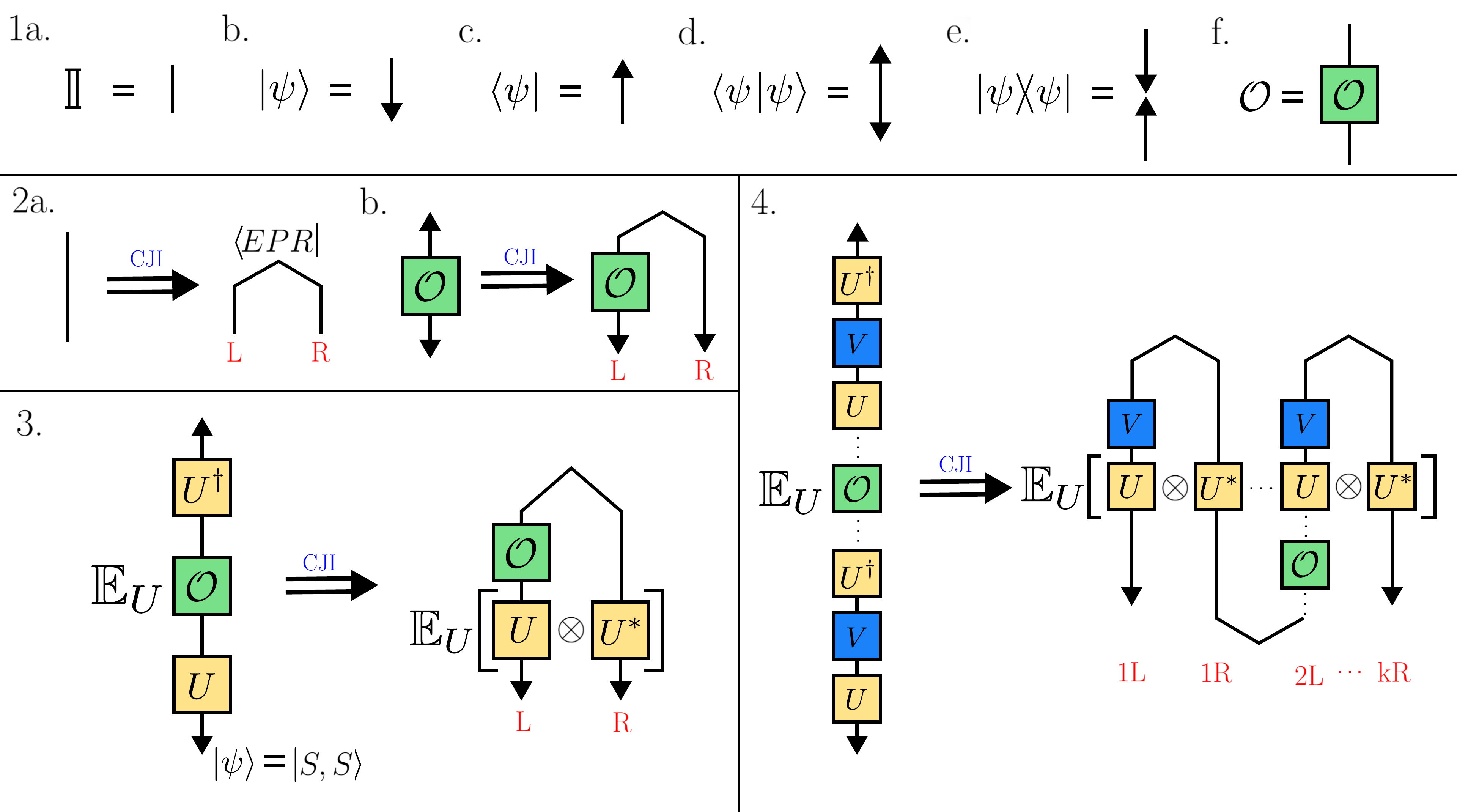}
    \caption{Visual representation of the Choi-Jamio\l kowski isomorphism (CJI) depicted using tensor-network notation. Notation is defined in panel 1 with: (1a) identity, (1b-c) states, (1d-e) products, and (1f) operators. The CJI acts turns the bras or kets that make up an operator into their counterpart. For instance (2a) shows the transformation on the identity for a spin-1/2 system, which maps $\ket{0}\bra{0} + \ket{1}\bra{1} \rightarrow \bra{0}_L \otimes \bra{0}_R + \bra{1}_L \otimes \bra{1}_R = \bra{EPR}_{LR}$. (2b) shows how expectation values of operators on states transform under the isomorphism: $\bra{\psi} \mathcal{O} \ket{\psi} \rightarrow \bra{EPR}_{LR} (\mathcal{O}_L \otimes \mathbb{I}_R) \ket{\psi}_L \otimes \ket{\psi}_R$. (3) When taking expectation values averaged over Haar random unitaries $U$, we can use the CJI to isolate the $U$ operators, absorbing any other operators into our boundary conditions. (4) shows a more complicated version of the same idea. Here, $k$ is the number of pairs of replicas of $U$.}
    \label{fig:Choi-Jam}
\end{figure}

\begin{figure}
    \centering
    \includegraphics[width=0.7\linewidth]{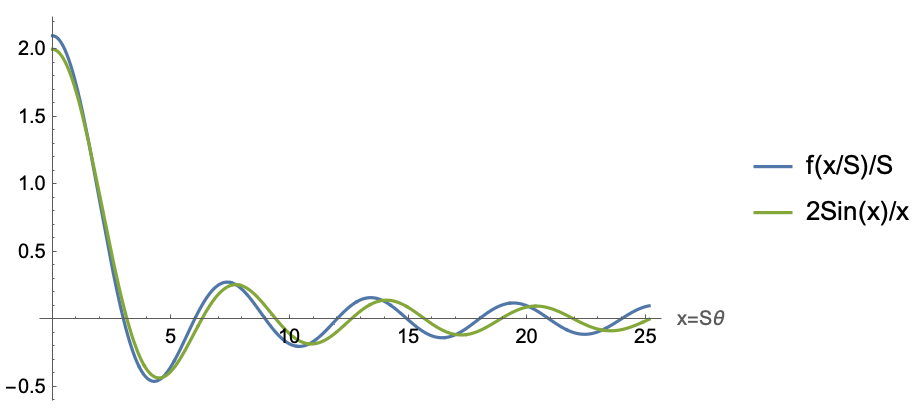}
    \caption{The function $f(\theta)/S$ plotted versus $x = S \theta$ for $S = 10$ (blue) and its asymptotic large-$S$ behavior (green).}
    \label{fig:fthetafunc}
\end{figure}

Next we consider the butterfly echo protocol, which prepares the final state
\begin{equation}
    \ket{\chi} = U^{\dagger} e^{-i \sn \theta} U \ket{S,S},
\end{equation}
and we are interested in computing ensemble-averaged expectation values
\begin{equation}
    \overline{\bra{\chi} \mathcal{O} \ket{\chi}} = \mathbb{E}_{U \sim \mathcal{H}} \left[ \bra{S,S} U^{\dagger} e^{i \sn \theta} U \mathcal{O} U^{\dagger} e^{-i \sn \theta} U \ket{S,S} \right]
\end{equation}
in this final state.
Such quantities are categorized as second-moment ($k = 2$) quantities because they involve two pairs of forward and backward evolutions $U, U^{\dagger}$. Using the Choi-Jamio\l kowski isomorphism (see Figure \ref{fig:Choi-Jam}) and the Weingarten calculus, we find that expectation values are given by
\begin{align}
    \overline{\bra{\chi} \mathcal{O} \ket{\chi}} &= \frac{1}{D^2-1} \left[ f^2(\theta) \bra{S,S} \mathcal{O} \ket{S,S} + D \tr{\mathcal{O}} \right] \nonumber \\
    & - \frac{1}{D(D^2-1)} \left[ f^2(\theta) \tr{\mathcal{O}} + D \bra{S,S} \mathcal{O} \ket{S,S} \right] \nonumber \\
    &= \tr{\mathcal{O}} / D + \frac{1}{D^2-1} \left( f^2(\theta) -1 \right) \left( \bra{S,S} \mathcal{O} \ket{S,S} - \tr{\mathcal{O}} / D \right)
    \label{eq:k2expval}
\end{align}
where
\begin{equation}
    f(\theta) = \tr{e^{-i \sn \theta}} = \sum_{m = -S}^S e^{-i m \theta} = \cos{(S \theta)} + \cot{(\theta / 2)} \sin{(S \theta)}.
\end{equation}
At large $S$ and fixed $x = S \theta$, this function asymptotes to the sinc function:
\begin{equation}
    \lim_{S \rightarrow \infty} f(x/S) / S = 2 \frac{\sin x}{x} \equiv 2 \ \mathrm{sinc}(S \theta)
\end{equation}
as shown in Fig. \ref{fig:fthetafunc}.

\begin{figure}
    \centering
    \includegraphics[width=0.7\linewidth]{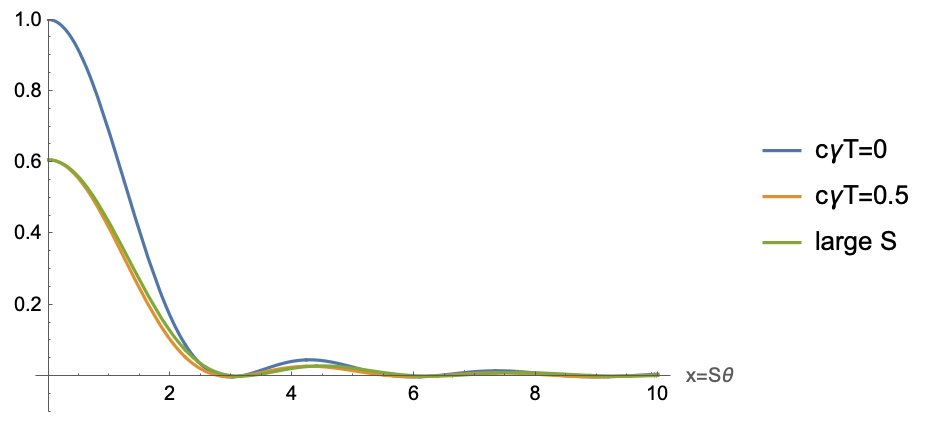}
    \caption{The expected signal $\langle S_z \rangle/S = \overline{\bra{\chi} S_z \ket{\chi}} / S$ as a function of $x = S \theta$ for $S = 10$ and $c \gamma T = 0$ (blue) and $c \gamma T = 0.5$ (orange). We also plot the asymptotic large-$S$ behavior with $c \gamma T = 0.5$ (green).}
    \label{fig:sztheta}
\end{figure}

\begin{figure}
    \centering
    \includegraphics[width=0.7\linewidth]{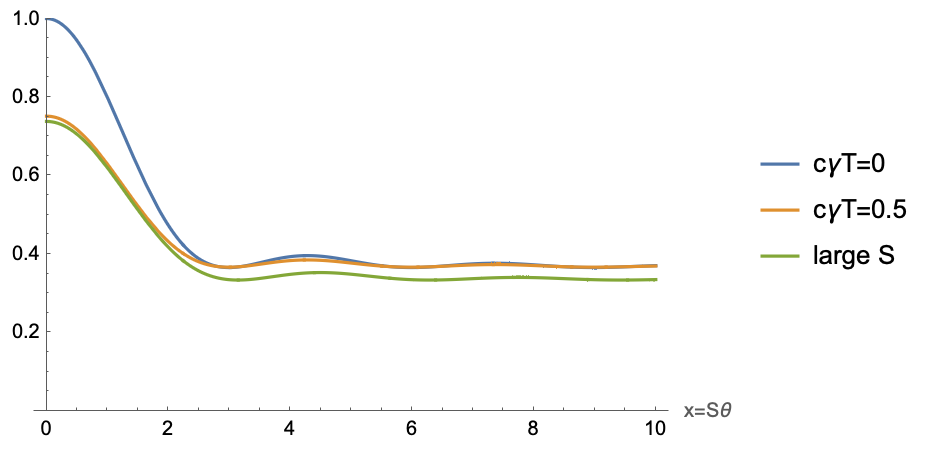}
    \caption{The expectation value $\langle S_z^2 \rangle/S^2 = \overline{\bra{\chi} S_z^2 \ket{\chi}} / S^2$ as a function of $x = S \theta$ for $S = 10$ and $c \gamma T = 0$ (blue) and $c \gamma T = 0.5$ (orange). We also plot the asymptotic large-$S$ behavior with $c \gamma T = 0.5$ (green).}
    \label{fig:sz2theta}
\end{figure}

\begin{figure}
    \centering
    \includegraphics[width=0.7\linewidth]{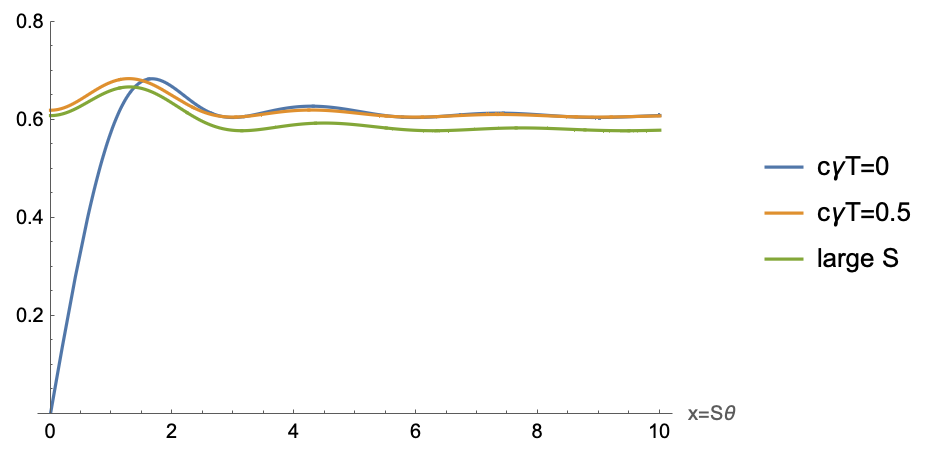}
    \caption{Quantum fluctuations $\Delta S_z / S$ as a function of $x = S \theta$ for $S = 10$ and $c \gamma T = 0$ (blue) and $c \gamma T = 0.5$ (orange). We also plot the asymptotic large-$S$ behavior with $c \gamma T = 0.5$ (green). Notice that the depolarizing noise has completely erased the linear behavior at small $x$ observed in the clean case.}
    \label{fig:szflucs}
\end{figure}

\begin{figure}
    \centering
    \includegraphics[width=0.7\linewidth]{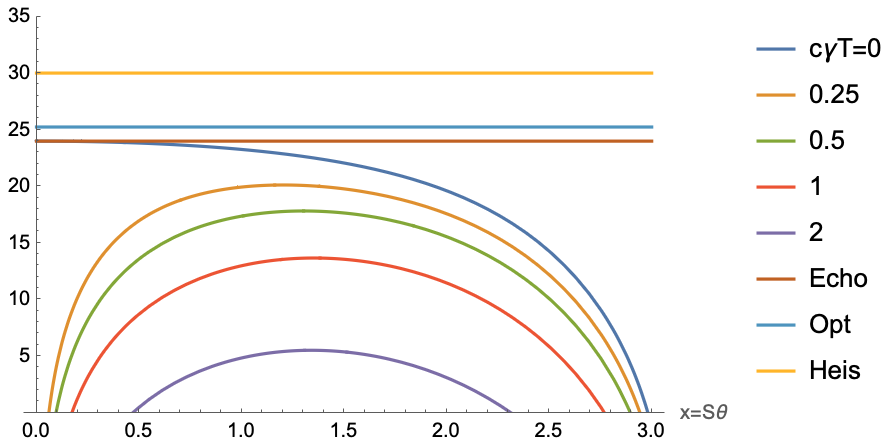}
    \caption{Metrological gain $G$ as a function of $x = S \theta$ for $N = 10^3$ and depolarizing noise $c \gamma T = 0,0.25,0.5,1,2$, compared to the maximum gain for the echo protocol (brown), the optimal gain for random Dicke states (light blue), and the Heisenberg limit (yellow).}
    \label{fig:metrgain}
\end{figure}

\begin{figure}
    \centering
    \includegraphics[width=0.7\linewidth]{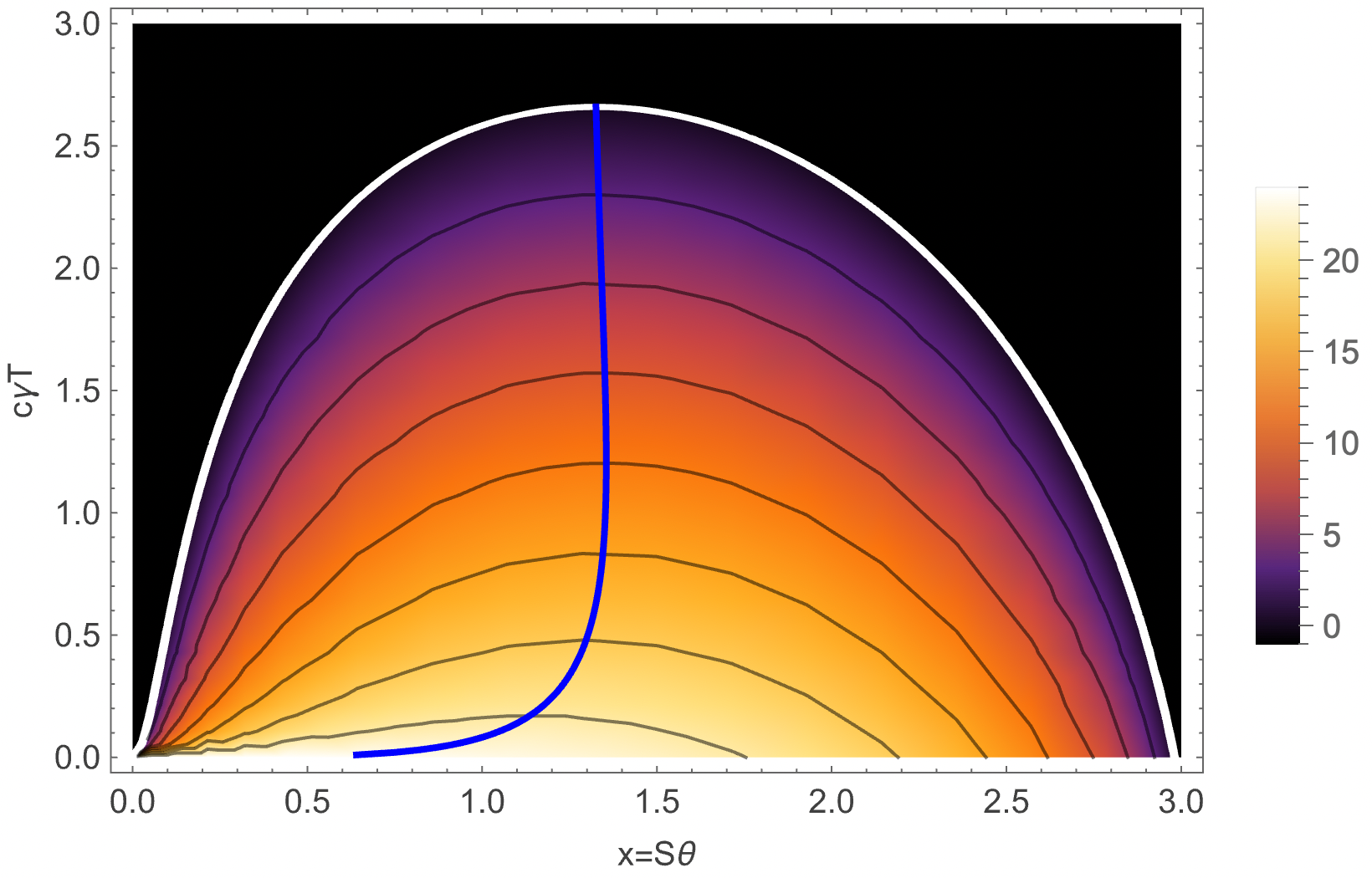}
    \caption{Metrological gain $G$ as a function of $x = S \theta$ and $c \gamma T$ for $N = 10^3$. Black contours are drawn at 3dB intervals and the blue curve shows the optimal sensing angle $x^*$ as a function of noise $c\gamma T$. The black region indicates parameter regimes with no metrological gain and the white boundary illustrates the phenomenon of bandwidth narrowing.}
    \label{fig:metrgaindensitycontour}
\end{figure}

We can use this expression to compute the metrological signal, which is governed by the expectation value of $S_z$:
\begin{equation}
    \overline{\bra{\chi} S_z \ket{\chi}} = \frac{S}{(2S+1)^2-1} \left[ f^2(\theta)-1 \right].
\end{equation}
At large $S$ and fixed $x = S \theta$ this asymptotes to:
\begin{equation}
    \lim_{S \rightarrow \infty} \overline{\bra{\chi} S_z \ket{\chi}} / S = \mathrm{sinc}^2 (S \theta)
\end{equation}
as shown in Fig. \ref{fig:sztheta}.
We can also use it to compute the noise in the signal caused by quantum fluctuations, which is governed by the expectation value of $S_z^2$:
\begin{equation}
    \overline{\bra{\chi} S_z^2 \ket{\chi}} = \frac{1}{3} S (S+1) + \frac{(2 S-1)}{12 (S+1)} \left[ f^2(\theta) -1 \right].
\end{equation}
At large $S$ and fixed $x = S \theta$ this asymptotes to:
\begin{equation}
    \lim_{S \rightarrow \infty} \overline{\bra{\chi} S_z^2 \ket{\chi}} / S^2 = \frac{1}{3} \left[1 + 2 \ \mathrm{sinc}^2 (S \theta) \right]
    \label{eq:sz2asympt}
\end{equation}
as shown in Fig. \ref{fig:sz2theta}.
Putting this all together, we obtain an expression for the quantum fluctuations
\begin{equation}
    \Delta S_z = \sqrt{\overline{\bra{\chi} S_z^2 \ket{\chi}} - \overline{\bra{\chi} S_z \ket{\chi}}^2}
\end{equation}
which has asymptotic behavior
\begin{equation}
    \lim_{S \rightarrow \infty} \Delta S_z / S = \frac{1}{\sqrt{3}} \sqrt{ 1+ 2 \ \mathrm{sinc}^2(S \theta) - 3 \ \mathrm{sinc}^4(S \theta) }
\end{equation}
as shown in Fig. \ref{fig:szflucs}.
Finally, we can combine these results to compute the angular sensitivity
\begin{align}
    \frac{1}{\Delta \theta} &= \frac{\partial \left(S - \overline{\bra{\chi} S_z \ket{\chi}} \right) / \partial \theta}{\Delta S_z} \nonumber \\
    &\approx N \ \frac{\sqrt{3}}{x} \frac{\sinc(x) \left( \sinc(x) - \cos(x) \right)}{\sqrt{1+2 \sinc^2(x) -3 \sinc^4(x)}} + \mathcal{O}(1) \nonumber \\
    &\approx \frac{N}{2} \left( 1 - 3 x^2 / 40 + \mathcal{O}(x^4) \right) + \mathcal{O}(1)
    \label{eq:angsensclean}
\end{align}
where we have truncated the expression to the leading term in the large $N$ limit.
We see that the angular sensitivity is sub-optimal near $x = 0$ by a factor of $\sqrt{3}/2 \approx 0.866$ compared to the angular sensitivity of $\Delta \theta^{-1} \approx N / \sqrt{3}$ that we expect from the Cramer-Rao bound and the quantum Fisher information of a random Dicke state. We plot the resulting metrological gain
\begin{equation}
    G = 10 \log_{10} \left( \frac{1}{N \Delta \theta^2} \right)
\end{equation}
in Figs. \ref{fig:metrgain} and \ref{fig:metrgaindensitycontour}.

\subsection{Bounding the Fluctuations: Second, Third, and Fourth Moment Analysis}

Crucially, the random choice of unitary operators $U$ will lead to fluctuations in the QFI $\mathcal{F}_{\hat{n}}(\theta)$ and the metrological signal $\langle S_z \rangle$. One might reasonably be concerned that this additional source of noise will overwhelm the metrological signal but we show here that these fluctuations are strictly smaller than the inherent quantum fluctuations. These fluctuations come in two flavors: for a fixed rotation axis $\hat{n}$ we expect to see shot-to-shot fluctuations in the QFI and the metrological signal as we vary the preparation unitary $U$. For the QFI, these fluctuations are captured by the variance
\begin{equation}
    \mathrm{Var}_U\left[\mathcal{F}_{\hat{n}}(\theta) \right] = \overline{\mathcal{F}^2} -  \overline{\mathcal{F}}^2 = \overline{\mathrm{Var}[\sn]^2} - \overline{\mathrm{Var}[\sn]}^2
\end{equation}
where the first term involves contributions from moments $k=2,3,4$ and the second term was computed earlier. For the metrological signal, these fluctuations are captured by the variance
\begin{equation}
    \mathrm{Var}_U\left[ \bra{\chi} S_z \ket{\chi} \right] = \overline{\bra{\chi} S_z \ket{\chi}^2} - \left( \overline{\bra{\chi} S_z \ket{\chi}} \right)^2
\end{equation}
where the first term contains contributions from the fourth moment $k = 4$ and the second term was computed earlier. Crucially, neither of these quantities depends on the rotation axis $\hat{n}$, so there is no need to perform additional averaging over the rotation axis.

In addition to shot-to-shot fluctuations, for a fixed unitary $U$ we expect to see anisotropic fluctuations in the signal as we vary the rotation axis $\hat{n}$. For the QFI, these fluctuations are captured by the variance
\begin{equation}
    \overline{\mathrm{Var}_{\hat{n}}\left[\mathcal{F}_{\hat{n}}(\theta) \right]} = \overline{\mathbb{E}_{\hat{n}} \left[\mathcal{F}_{\hat{n}}^2\right]} -  \overline{\mathbb{E}_{\hat{n}} \left[\mathcal{F}_{\hat{n}}\right]^2} = \overline{\mathcal{F}^2} - \mathbb{E}_{\hat{n},\hat{m}} \left[ \ \overline{\mathcal{F}_{\hat{n}} \mathcal{F}_{\hat{m}}} \ \right]
\end{equation}
where we have exchanged the order of the averages over $\hat{n},U$ and dropped the $\hat{n}$ average from the first term since the quantity does not depend on the rotation axis. The first term is the same as appeared above. For the second term we have explicitly written out the square of averages to emphasize that two different rotation axes $\hat{n},\hat{m}$ appear in the calculation. The second term involves moments $k = 2,3,4$. For the metrological signal, these anisotropic fluctuations are captured by the variance
\begin{align}
    \overline{\mathrm{Var}_{\hat{n}}\left[ \bra{\chi} S_z \ket{\chi} \right]} &= \overline{\mathbb{E}_{\hat{n}} \left[ \bra{\chi} S_z \ket{\chi}_{\hat{n}}^2 \right]} - \overline{ \mathbb{E}_{\hat{n}} \left[ \bra{\chi} S_z \ket{\chi}_{\hat{n}} \right]^2} \nonumber \\
    &= \overline{\bra{\chi} S_z \ket{\chi}^2} - \mathbb{E}_{\hat{n},\hat{m}} \left[ \ \overline{  \bra{\chi} S_z \ket{\chi}_{\hat{n}} \bra{\chi} S_z \ket{\chi}_{\hat{m}}} \ \right]
    \label{eq:signalanisotr}
\end{align}
where the first term is the same as appeared above, and the second term involves a fourth moment $k = 4$ calculation.

In what follows we will use the Weingarten calculus to compute the leading order contributions in the large-$S$ limit. Because the calculations of the various quantities of interest involve similar steps, we only discuss the calculation of the quantity
\begin{equation}
    Q_{\hat{n},\hat{m}} = \overline{  \bra{\chi} S_z \ket{\chi}_{\hat{n}} \bra{\chi} S_z \ket{\chi}_{\hat{m}}}
\end{equation}
in detail, and simply report the results for the other quantities above. Using the Choi-Jamiolkowski isomorphism, we may express the quantity $Q_{\hat{n},\hat{m}}$ in terms of $2k$ replicas labeled by the indices $r = 1,2,\ldots,k$ and $a = L,R$ where $L$ indicates a forward-time replica $U$ and $R$ indicates a backward-time replica $U^{\mathcal{T}}$. Using the Weingarten calculus, the expectation value over the Haar ensemble yields a sum over all possible pairings of forward $U$ and time-reversed $U^{\mathcal{T}}$ replicas, where each term in the sum is weighted by the Weingarten functions $\mathrm{Wg}(\sigma \tau^{-1}, D)$ where $D=2S+1$ is the Hilbert space dimension \cite{collins2022weingarten}. For the present case we have $k = 4$, and $\sigma,\tau$ are elements of the symmetric group $S_4$ that parameterize the possible replica pairings. There are $|S_4| = 4! = 24$ possible elements, leading to a total of $(4!)^2 = 576$ terms, which we label by the tuple $[\sigma, \tau]$. Fortunately, most of these terms are subleading in the limit of large $S$, and our task in the following is to isolate only the leading-order terms and ignore the rest.

The leading term is the trivial pairing $[\sigma,\tau] = [\mathbb{I},\mathbb{I}]$ where $\sigma = \tau = \mathbb{I} \equiv \mathbb{I}^4$ is the identity element, and the superscript reminds us that we are working with $k = 4$ and permutation elements $\sigma,\tau \in S_4$. This yields the leading $\mathcal{O}(S^2)$ contribution
\begin{align}
    &\mathrm{Wg}(\mathbb{I}^4,D) \bra{S,S} S_z \ket{S,S}^2 \mathrm{Tr}^2 \left[ e^{-i S_{\hat{n}} \theta} \right] \mathrm{Tr}^2 \left[ e^{-i S_{\hat{m}} \theta} \right] \nonumber \\
    = &\mathrm{Wg}(\mathbb{I}^4,D) S^2 f^4(\theta) \quad \quad \quad \quad [\mathbb{I},\mathbb{I}]
    \label{eq:k4leading}
\end{align}
where the dependence on $\hat{n},\hat{m}$ disappears because the trace is axis-independent. Next we systematically identify all subleading terms of order $\mathcal{O}(S)$. There are two classes of terms that can appear: diagonal terms $[\sigma,\sigma]$, which always yield the leading-order Weingarten function $\mathrm{Wg}(\sigma \sigma^{-1},D) = \mathrm{Wg}(\mathbb{I}^4,D)$; and off-diagonal terms $[\sigma,\tau]$ with $\sigma \neq \tau$, which yield subleading Weingarten functions $\mathrm{Wg}(\sigma \tau^{-1},D)$. We start with the diagonal terms; a direct search through all $4! = 24$ possibilities yields the following 3 terms of order $\mathcal{O}(S)$:
\begin{align}
    &2 \mathrm{Wg}(\mathbb{I}^4,D) S^2 f^2(\theta) f(\mu_+(\theta,\hat{n}\cdot\hat{m})) \quad \quad \quad \quad [(2 \ 3),(2 \ 3)] \ \mathrm{and} \ [(1 \ 4),(1 \ 4)] \nonumber \\
    &2 \mathrm{Wg}(\mathbb{I}^4,D) S^2 f^2(\theta) f(\mu_-(\theta,\hat{n}\cdot\hat{m})) \quad \quad \quad \quad [(1 \ 3),(1 \ 3)] \ \mathrm{and} \ [(2 \ 4),(2 \ 4)] \nonumber \\
    &\mathrm{Wg}(\mathbb{I}^4,D) \frac{S(S+1)}{3} D f^2(\mu_+(\theta,\hat{n}\cdot\hat{m})) \quad \quad \quad \quad [(2 \ 3)(1 \ 4),(2 \ 3)(1 \ 4)]
    \label{eq:k4suba}
\end{align}
where $\sigma = (r \ s)$ is the transposition element that swaps replicas $r,s$. The transpositions $(1 \ 2)$ and $(3 \ 4)$ are absent from this list because the diagrams vanish exactly due to the fact that $\tr{S_z} = 0$. The angles $\mu_{\pm}$ originate in terms of the form
\begin{equation}
    \mathrm{Tr}\left[ e^{-i S_{\hat{n}} \theta} e^{\pm i S_{\hat{m}} \theta} \right] = f(\mu_{\pm}(\theta,\hat{n}\cdot\hat{m}))
\end{equation}
where the angles $\mu_{\pm}$ are given by the Rodrigues rotation formula
\begin{equation}
    \cos \frac{\mu_{\pm}}{2} = \cos^2 \frac{\theta}{2} \pm \left( \hat{n} \cdot \hat{m} \right) \sin^2 \frac{\theta}{2}.
\end{equation}
Note that the resulting expressions depend only on the rotation-invariant dot product $\hat{n} \cdot \hat{m}$, which makes sense because Haar-random averages are axis-independent.

Next we consider off-diagonal terms $[\sigma,\tau]$ with $\sigma \neq \tau$. To systematically identify the dominant terms, we take advantage of the fact that Weingarten functions $\mathrm{Wg}(\sigma,D)$ are reduced by factors of $1/D$ for each additional transposition that appears in the transposition decomposition of $\sigma$. For example, $\mathrm{Wg}(\mathbb{I}^4,D) \sim \mathcal{O}(D^{-4})$, whereas $\mathrm{Wg}(2 \mathbb{I}^2,D) \sim \mathcal{O}(D^{-5})$, where the notation $2 \mathbb{I}^2$ indicates a single transposition of 2 elements while the other 2 elements are left unchanged. Hence, to conduct our search we first look for pairs of permutations $\sigma, \tau$ that differ by a single transposition. This yields the following 2 terms of order $\mathcal{O}(S)$:
\begin{align}
    &4 \mathrm{Wg}(2\mathbb{I}^2,D) S^2 f^4(\theta) \quad \quad \quad \quad \quad \quad \quad \quad \quad \quad \quad \quad \quad \quad [\mathbb{I},\sigma] \ \mathrm{for} \ \sigma = (2 \ 3), (1 \ 4), (2 \ 4), (1 \ 3) \nonumber \\
    &2 \mathrm{Wg}(2\mathbb{I}^2,D) \frac{S(S+1)}{3} D f^2(\theta) f(\mu_+(\theta,\hat{n}\cdot\hat{m})) \quad \quad \quad \quad [\sigma,(2 \ 3)(1 \ 4)] \ \mathrm{for} \ \sigma = (2 \ 3), (1 \ 4)
    \label{eq:k4subb}
\end{align}
where the transpositions $\sigma = (1 \ 2), (3 \ 4)$ do not appear in the first line because $\tr{S_z} = 0$. Finally, we find one remaining $\mathcal{O}(S)$ term generated by 2 pairs of transpositions:
\begin{equation}
    \mathrm{Wg}(2^2,D) \frac{S(S+1)}{3} D f^4(\theta) \quad \quad \quad \quad [\mathbb{I},(2 \ 3)(1 \ 4)]
    \label{eq:k4subc}
\end{equation}
Summing together all seven terms from Eqs. \eqref{eq:k4leading}, \eqref{eq:k4suba}, \eqref{eq:k4subb}, and \eqref{eq:k4subc}, we obtain an expression for $Q_{\hat{n},\hat{m}}$ that is accurate to order $\mathcal{O}(S)$ for large $S$.

Finally we use these results to compute the leading order anisotropic variance in the metrological signal Eq. \eqref{eq:signalanisotr}. To do so we must take the expectation value over rotation axes $\hat{n},\hat{m}$, which involves dealing with the functions $g(\mu_{\pm}) \equiv f(\mu_{\pm}(\theta,\hat{n} \cdot \hat{m})) / (2S+1) $. Expanding Eq. \eqref{eq:signalanisotr} to leading order in $S$ we obtain
\begin{equation}
    \frac{S}{6} \mathbb{E}_{\hat{n},\hat{m}} \left[ 1 - g^2(\mu_+) + 4 g^2(\theta) \left( 1 - g(\mu_+) + \frac{3}{2} g(2\theta) - \frac{3}{2} g(\mu_-) \right) \right]
\end{equation}
and expanding to leading order in $x = S \theta$ we obtain
\begin{equation}
    \mathbb{E}_{\hat{n},\hat{m}} \left[ \frac{23}{270} \left( 1 - (\hat{n} \cdot \hat{m})^2 \right) S x^4 + S \mathcal{O}(x^6) \right]
\end{equation}
which yields a variance of order
\begin{equation}
    \overline{\mathrm{Var}_{\hat{n}}\left[ \bra{\chi} S_z \ket{\chi} \right]} = \frac{23}{324} S x^4 + S \mathcal{O} (x^6)
\end{equation}
at lowest order in $x$ and $1/S$. The same techniques can be used to compute the remaining variances above, which yield:
\begin{equation}
    \mathrm{Var}_U\left[ \bra{\chi} S_z \ket{\chi} \right] = \frac{17}{270} S x^4 + S \mathcal{O} (x^6)
\end{equation}
\begin{equation}
    \mathrm{Var}_U\left[\mathcal{F}_{\hat{n}}(\theta) \right] = \frac{4}{45} N^3 + \mathcal{O}(N^2)
\end{equation}
All of these variances are subleading compared to the dominant quantum fluctuations.

\section{Perturbation Theory for Noisy Echo Protocol}
\label{app:noisypert}

In this section we consider the effects of decoherence applied during state preparation and time-reversal. We assume an isotropic depolarizing channel described by the Lindblad master equation with jump operators $L_{j} = \sqrt{\gamma} S_j$ for $j = x,y,z$ where $\gamma$ is the depolarizing rate. The Choi-Jamio\l kowski isomorphism combined with disorder averaging \cite{choi1975completely,jamiolkowski1972linear,bentsen2021measurement} yield an effective Hamiltonian
\begin{equation}
    H_{\mathrm{eff},\gamma} = \heff + \gamma V
\end{equation}
where $\heff$ is the clean (noiseless) effective Hamiltonian obtained from the random OAT (or Brownian) model, and
\begin{equation}
    V = \sum_{r=1}^k \left( S(S+1) + \vec{S}_{rL} \cdot \vec{S}_{rR} \right)
\end{equation}
is the effective Hamiltonian generated by the depolarizing channel. Here we assume a perturbatively small depolarizing rate $\gamma \ll J$ and sufficiently large circuit depth $JT \gg 1$ such that the excited states of $\heff$ are not relevant to the long-time dynamics. In this case we can restrict our attention to the ground subspace and perform standard degenerate perturbation theory to compute the shifted ground states and their energies. The case $k = 1$ is trivial: in this case $\heff$ has a unique ground state $\ket{\Omega}$ which is also an exact eigenstate of $V$ with vanishing eigenvalue, so the ground state $\ket{\Omega}$ is not shifted by $V$.

For $k = 2$ we have a 2-dimensional ground subspace spanned by the states $\ket{\alpha},\ket{\beta}$ corresponding to the `ladder' and `crossed' saddle points, respectively. Note that these states are normalized but are \emph{not orthogonal}:
\begin{equation}
    \braket{\alpha}{\beta} = \frac{1}{D}
\end{equation}
To obtain an orthonormal basis for the ground subspace, we may use the standard Gram-Schmidt procedure to obtain
\begin{equation}
    \ket{\mu} = \frac{D \ket{\beta} - \ket{\alpha}}{\sqrt{D^2-1}}
\end{equation}
which guarantees $\braket{\mu}{\mu} = 1$ and $\braket{\mu}{\alpha} = 0$. We now apply standard degenerate perturbation theory, which involves computing matrix elements of $V$:
\begin{align}
    \bra{\alpha} V \ket{\alpha} &= 0 \nonumber \\
    \bra{\mu} V \ket{\mu} &= 2 S(S+1) \frac{D^2}{D^2-1} \equiv c \nonumber \\
    \bra{\alpha} V \ket{\mu} &= 0 \nonumber \\
\end{align}
Because the off-diagonal elements vanish, the eigenstates $\ket{\alpha},\ket{\mu}$ themselves are not shifted as a result of the perturbation $V$. The only change is the first-order shift $c \gamma$ in the energy of $\ket{\mu}$.

These results allow us to generalize Eq. \eqref{eq:k2expval} in the presence of dissipation. In the deep-circuit limit $JT \rightarrow \infty$ but where we keep $\gamma T$ fixed, we obtain
\begin{align}
    \overline{\bra{\chi} \mathcal{O} \ket{\chi}} &= \frac{D^2-e^{-c \gamma T}}{D^2-1} \tr{\mathcal{O}} / D + \frac{e^{-c \gamma T}}{D^2-1} \left( f^2(\theta) -1 \right) \left( \bra{S,S} \mathcal{O} \ket{S,S} - \tr{\mathcal{O}} / D \right)
    \label{eq:k2expvalnoisy}
\end{align}
which we can immediately use to compute the metrological signal
\begin{equation}
    \overline{\bra{\chi} S_z \ket{\chi}} = \frac{S e^{-c \gamma T}}{D^2-1} \left[ f^2(\theta)-1 \right].
\end{equation}
which decays exponentially with the circuit depth. Similarly, we can compute the second moment
\begin{equation}
    \overline{\bra{\chi} S_z^2 \ket{\chi}} = \frac{1}{3} S (S+1) \frac{D^2-e^{-c \gamma T}}{D^2-1} + e^{-c \gamma T} \frac{(2 S-1)}{12 (S+1)} \left[ f^2(\theta) -1 \right].
\end{equation}
At large $S$ and fixed $x = S \theta$ this asymptotes to:
\begin{equation}
    \lim_{S \rightarrow \infty} \overline{\bra{\chi} S_z^2 \ket{\chi}} / S^2 = \frac{1}{3} \left[1 + e^{-c \gamma T} 2 \ \mathrm{sinc}^2 (S \theta) \right]
\end{equation}
which generalizes Eq. \eqref{eq:sz2asympt}. Combining these results we obtain the angular sensitivity
\begin{align}
    \frac{1}{\Delta \theta} &= \frac{\partial \left(S - \overline{\bra{\chi} S_z \ket{\chi}} \right) / \partial \theta}{\Delta S_z} \nonumber \\
    &\approx N \ \frac{\sqrt{3}}{x} \frac{e^{-c \gamma T} \sinc(x) \left( \sinc(x) - \cos(x) \right)}{\sqrt{1+2 \sinc^2(x) e^{-c \gamma T} -3 \sinc^4(x) e^{-2 c \gamma T}}} + \mathcal{O}(1)
    \label{eq:angsensnoisy}
\end{align}
where we have truncated the expression to the leading term in the large $N$ limit. This expression generalizes Eq. \eqref{eq:angsensclean} to the noisy case. Eq. \eqref{eq:angsensnoisy} is one of our key technical contributions from the analytic calculations. We anticipate that at sufficiently large $\gamma$ there is a phase transition in $H_{\mathrm{eff},\gamma}$ to noise-dominated dynamics that are useless for metrology. We leave the study of this phase transition to future work.

\section{Derivation of Effective Hamiltonian Spectrum} \label{app:HeffDerivation}

In this section, we analyze a Brownian circuit model for generating random symmetric probe states and show that it is equivalent to the random one axis twisting (ROAT) model in the limit of infinitesimal twists. Using the Choi-Jamio\l kowski isomorphism and disorder averaging \cite{choi1975completely,jamiolkowski1972linear,bentsen2021measurement} we can describe the circuit dynamics in terms of an effective Hamiltonian $\heff^{(k)}$ where $2k$ is the number of replicas (Fig. \ref{fig:Choi-Jam}). Analysis of the spectrum of this effective Hamiltonian shows an energy gap that is constant in spin size $S$, indicating that the timescale required to reach the same metrological usefulness as a Haar-random probe state is constant in system size.

We first study a Brownian model and later show that the dynamics of this model are identical to a random one-axis twisting model in the limit of infinitesimally small twisting times. The dynamics in the Brownian model consist of a series of short pulses, governed by a unitary operator
\begin{equation}\label{eq:HBrownian}
    U_B := \prod_t U_{B,t} = \prod_t \exp{(-iH_{B}(t) \ dt)}, ~ ~ ~ ~ H_{B}(t) := \sum_{\alpha\beta} J^{\alpha\beta}(t) S^\alpha S^\beta
\end{equation}
where $S^{\alpha}$ are angular momentum operators with $\alpha,\beta = x,y,z$ and total spin $S$. Here $J^{\alpha\beta}(t)$ are Brownian random coupling coefficients, which are drawn from a Gaussian distribution with zero mean and variance

\begin{equation} \label{eq:BrownianVariance}
    \mathbb{E}_J[J^{\alpha\beta}(t) J^{\alpha'\beta'}(t')] =  \frac{J\delta^{tt'}}{S^2 dt}(\delta^{\alpha\alpha'}\delta^{\beta\beta'}+\delta^{\alpha\beta'}\delta^{\alpha'\beta})
\end{equation}
where $\mathbb{E}_J$ denotes the ensemble average over couplings, and the coupling strength $J$ sets the overall energy scale.

For simplicity we first use this model to study the mean quantum Fisher information (QFI) for the probe state $\ket{\psi} = U_B \ket{S,S}$, which is given by
\begin{align}
    \overline{\mathcal{F}_{\hat{n}}(\theta)} &= 4\mathbb{E}_J \left[ \mathrm{Var}\left( S_{\hat{n}} \right) \right] \nonumber \\
    &= 4\mathbb{E}_J \left[ \bra{S,S} U_B^{\dagger} \left( S_{\hat{n}} \right)^2 U_B \ket{S,S} - \bra{S,S} U_B^{\dagger} S_{\hat{n}} U_B \ket{S,S}^2 \right].
\end{align}
For the moment we focus on the first term, which involves two copies $U_B,U_B^{\dagger}$ of the evolution operator. Using standard techniques \cite{bentsen2021measurement}, we apply the Choi–Jamiołkowski isomorphism (channel-state duality) along with the time-reversal operator $\mathcal{T}$ to express this quantity in terms of the expectation value on two \emph{replicas}:
\begin{equation}
    \mathbb{E}_J\left[U_{B}\otimes U_{B}^{\mathcal{T}}\right] = \prod_t \mathbb{E}_J\left[U_{B,t}\otimes U_{B,t}^{\mathcal{T}}\right]
    \label{eq:Btexpval}
\end{equation}
where the absence of correlations in time allows us to take the ensemble average over each timestep $t$ independently. Via a Taylor expansion of \cref{eq:Btexpval} in powers of $J dt$ we obtain:

\begin{equation}\label{eq:expectationvalue}
    \begin{split}
        \mathbb{E}_J\left[U_t\otimes U_t^{\mathcal{T}}\right] &= \mathbb{I} - \frac{Jdt}{2S^2} \sum_{ab}\sum_{\alpha\beta}(-1)^{a + b} (S_{a}^\alpha S_{a}^\beta S_{b}^{\alpha} S_{b}^{\beta} + S_{a}^\alpha S_{a}^\beta S_{b}^{\beta} S_{b}^{\alpha}) + \mathcal{O}(dt^2)\\
        &= \mathbb{I} - \frac{Jdt}{2S^2} \left[\sum_a (2\vec{S}_a^4 - \vec{S}_a^2) + \sum_{a \neq b}(-1)^{a + b}(2(\vec{S}_{a}\cdot\vec{S}_{b})^2 + (\vec{S}_{a}\cdot\vec{S}_{b}))\right] + \mathcal{O}(dt^2)\\
    \end{split}
\end{equation}
where we use $a,b = L,R$ to index the two replicas for forward $U_B$ and time-reversed $U_B^{\mathcal{T}}$ evolution. The time-reversed unitary is computed through the insertion of the identity $(iY)(iY)^\dagger$, where $iY := \prod_{i=1}^N i\sigma_y$ for all individual spin-1/2 particles indexed by $i$ where $N$ is the total number of spins. Note that this insertion also affects the effective boundary conditions. Finally, we re-exponentiation the Taylor expansion in \cref{eq:expectationvalue} to get a partition function of the form $\mathbb{E}_U\left[(U_t\otimes U_t^{\mathcal{T}})\right] = \exp{(-\heff^{(1)}dt)}$ where $\heff^{(1)}$ is given by
\begin{equation}
    \heff^{(1)} = \frac{J}{2S^2} \left[\sum_a (2\vec{S}_a^4 - \vec{S}_a^2) + \sum_{a \neq b}(-1)^{a + b}(2(\vec{S}_{a}\cdot\vec{S}_{b})^2 + (\vec{S}_{a}\cdot\vec{S}_{b}))\right]
    \label{eq:Heff1}
\end{equation}
Although the above equation applies when there are only a single pair of replicas, quantities involving $k > 1$ pairs of unitary operators $U,U^{\dagger}$ the Choi–Jamiołkowski yields $2k$ replicas indexed by $a,b = L,R$ and $r,s = 1,2,\ldots,k$. A calculation similar to above leads to the following effective Hamiltonian
\begin{equation}
    \heff^{(k)} = \frac{J}{2S^2} \left[\sum_{ra} (2\vec{S}_{ra}^4 - \vec{S}_{ra}^2) + \sum_{ra \neq sb}(-1)^{a + b}(2(\vec{S}_{ra}\cdot\vec{S}_{sb})^2 + (\vec{S}_{ra}\cdot\vec{S}_{sb}))\right],
    \label{eq:Heffk}
\end{equation}
which generalizes Eq. \eqref{eq:Heff1}. The spectrum of this Hamiltonian, and the ground-state energy gap in particular, determines the timescales at which the random probe states become metrologically useful.

Before analyzing the effective Hamiltonian above, we show that this Brownian model is equivalent to the random one-axis twisting model considered in the main text in the limit of infinitesimal twisting pulses. In particular, the random OAT unitary is given by
\begin{equation}
    U_{\mathrm{R}} := \prod_t U_{\mathrm{R},t} = \prod_t \exp{(-iH_{\mathrm{R}}(t) \Delta t)}, ~ ~ ~ ~ H_{\mathrm{R}}(t) := \chi (\vec{S} \cdot \hat{m}_t)^2,
    \label{eq:HROAT}
\end{equation}
where $\chi$ is a coupling controlling the twisting strength and $\Delta t$ is the size of the timestep. At each timestep $t$ we choose a twisting axis $\hat{m}_t$ uniformly at random on the unit sphere. Similar to above, we map the problem onto a pair of replicas, take a Taylor expansion in the small parameter $\chi \Delta t S^2 \ll 1$, and perform the ensemble average over the twisting axes $\hat{m}_t$. Using the second- and fourth-moment expressions for ensemble averages over unit vectors
\begin{align}
    \int d^2 \hat{m} \ m_i m_j &= \frac{1}{3} \delta_{ij} \nonumber \\
    \int d^2 \hat{m} \ m_i m_j m_k m_{\ell}  &= \frac{1}{15} \left( \delta_{ij} \delta_{k \ell} +\delta_{ik} \delta_{j \ell} + \delta_{i \ell} \delta_{j k} \right)
    \label{eq:munitmoments}
\end{align}
we find an effective Hamiltonian
\begin{equation}
    H_{\mathrm{eff,R}}^{(1)} = \frac{\chi^2 S^2 \Delta t}{15J} \heff^{(1)} = \chi \left( \frac{\chi \Delta t}{30} \right) \left[\sum_a (2\vec{S}_a^4 - \vec{S}_a^2) + \sum_{a \neq b}(-1)^{a + b}(2(\vec{S}_{a}\cdot\vec{S}_{b})^2 + (\vec{S}_{a}\cdot\vec{S}_{b}))\right]
    \label{eq:Heff1ROAT}
\end{equation}
which matches the Brownian Hamiltonian up to an overall multiplicative factor.
A similar calculation holds for higher replica moments $k > 1$. Thus, conclusions drawn from analyzing the spectrum of the Brownian model (as done in the following section) apply directly to the random OAT model in the small twisting strength limit.

\begin{figure}[h]
    \centering
    \includegraphics[width=0.7\linewidth]{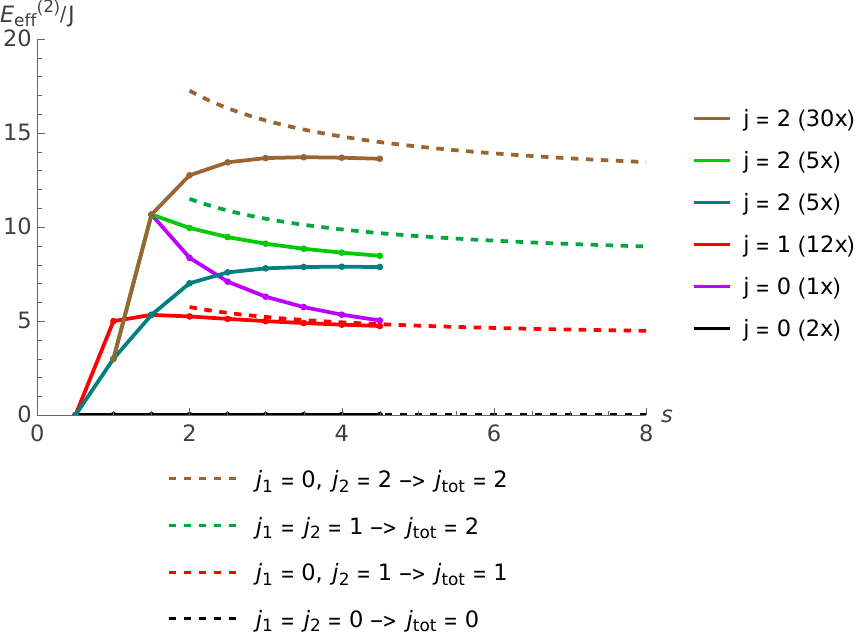}
    \caption{Effective energy spectrum of $H^{(2)}_{\mathrm{eff}}$ for $k = 2$ for different values of total spin $j$, where the degeneracies of each level are labelled in the legend -- for example, the ground states (black) are doubly degenerate (2x).  Analytic results are shown with dashed lines and numeric results are shown with solid lines.  Analytics are shown labeled for two different spin pairs, each with independent spin excitations, with both pairs being LR pairs. These also indicate combined angular momentum. The numerics are shown by their total angular momentum only, as specific pairings do not have good quantum numbers associated with them in the non-mean-field case. Therefore, to differentiate, these are also labeled by degeneracy. The analytics and numerics are color-coded to associate different levels together with the sole exception being the excited singlet state in the numerics, which comes from remnants of features lost in the mean-field approximation.}
    \label{fig:HeffSpectrum}
\end{figure}

\subsection{Analysis of $H_{\rm eff}^{(k)}$ Spectrum}
In this section we analyze the spectrum of the effective Hamiltonian of the Brownian circuit model described in the previous section. The eigenvalue gap of the effective Hamiltonian directly corresponds to the circuit depths needed to generate metrologically useful probe states via these dynamics. We find that the gap is independent of system size, corroborating our numerical evidence that random OAT generates metrologically useful states with a constant number of twists.

The spectrum of the $k = 1$ effective Hamiltonian is exactly solvable yielding energies 

\begin{equation}\label{eq:Eeff1}
    E_{\mathrm{eff}}^{(1)}(S,j) = \frac{J}{2S^2}j(j+1)(4S(S+1)-j(j+1)-1)
\end{equation}
 where $j$ is the total combined spin of both replicas, which can be any integer in the range $0 \leq j \leq 2S$. In the limit that $S\gg j$, $E_{\rm eff}^{(1)}(S,j) \rightarrow 2Jj(j+1)$, yielding an energy gap of $\Delta E_{\rm eff}^{(1)} = E_{\rm eff}^{(1)}(S,j=1) - E_{\rm eff}^{(1)}(S,j=0) \rightarrow 4J$.  Thus, our effective energy gap does not scale with $S$.

For $k > 1$, we apply mean field theory to \cref{eq:Heffk} to obtain the following effective mean-field Hamiltonian for arbitrary $k$:

\begin{equation}
    \begin{split}
        H_{\rm eff, mf}^{(k)} = \frac{J}{2S^2} [&2k(2S^2(S+1)^2 - S(S+1)) \\
        &- \sum_{ra < sb} (-1)^{a+b} (4G_{rasb}^2 -(\vec{J}_{rasb}^2 - 2S(S+1))(4G_{rasb} + 1))]
    \end{split}
\end{equation}
where $G_{rasb}$ is the mean-field condition for a given pairing of replica $ra$ to replica $sb$, and $\vec{J}_{rasb}$ is the corresponding combined angular momentum. This mean field effective Hamiltonian is valid for all $k$, but for the remainder of this section we restrict our attention to $k=2$.

For $k = 2$, limiting ourselves to just $rLsR$ average singlet pairings---i.e. setting $G_{rLsR}$ to $-S(S+1)$ and other $G_{rasb}$ to 0---we get a spectrum of the following form:
\begin{equation}
    E_{\mathrm{eff}}^{(k)} = \frac{J}{2S^2}(4S(S+1)-1) \sum_{rs} j_{rLsR}(j_{rLsR} + 1)
\end{equation}
which, again, yields an effective energy gap of $4J$ in the $S\gg j$ limit, leading us to the conclusion that the approach to randomness for $k=2$ occurs on constant timescales just like for the $k=1$ case. 

In \cref{fig:HeffSpectrum} we compare our analytical mean-field results at $k=2$ for the spectrum to numerical results found via exact diagonalization of \cref{eq:Heffk}. We observe that the low-lying spectrum found via numerics asymptotically approaches the mean field results from the effective Hamiltonian, although we have not fully understood the degeneracies in the spectrum of the finite-size numerics. However, as the eigenvalue gap between ground state and first excited state of the effective Hamiltonian is what we are after, the numerics give us confidence that our approximations are correct and that, indeed, the gap is independent of $S$.

\subsection{Convergence of Finite-Twist Random One-Axis Twist Model}

Whereas the above analysis focused on the spectrum of the Brownian circuit model, it is instructive to directly compare the numerical results presented in the main text to a random one-axis twisting model with finite twisting pulses. Such a model remains analytically tractable but is more relevant to experiments than the models featuring infinitesimally small pulses. Consider the random OAT unitary defined in Eq. \eqref{eq:HROAT} but now we take the twisting strength $\chi \Delta t$ to be finite instead of infinitesimal. In this case we need to be more careful about performing the Taylor expansion leading to the effective Hamiltonian. Our goal is to compute the ensemble average for each twisting pulse
\begin{equation}
    \mathbb{E}_{\hat{m}_t} \left[ U_{\mathrm{R},t} \otimes U_{\mathrm{R},t}^{\mathcal{T}} \right] = \int d^2 \hat{m}_t \ \exp \left[- i \chi \Delta t \left( (\vec{S}_L \cdot \hat{m}_t)^2 -  (\vec{S}_R \cdot \hat{m}_t)^2 \right) \right].
\end{equation}
We now Taylor-expand the exponential but take extra care to make sure that the higher-order terms can be ignored. The first-order term
\begin{equation}
    - i \chi \Delta t \int d^2 \hat{m}_t \left( (\vec{S}_L \cdot \hat{m}_t)^2 -  (\vec{S}_R \cdot \hat{m}_t)^2 \right) = 0
\end{equation}
vanishes exactly using the second-moment identity in Eq. \eqref{eq:munitmoments}. The second-order term is
\begin{equation}
    - \frac{1}{2} (\chi \Delta t)^2 \int d^2 \hat{m}_t \left( (\vec{S}_L \cdot \hat{m}_t)^2 -  (\vec{S}_R \cdot \hat{m}_t)^2 \right)^2 = - H_{\mathrm{eff,R}}^{(1)} \Delta t
\end{equation}
using the fourth-moment identity in Eq. \eqref{eq:munitmoments}, where $H_{\mathrm{eff,R}}^{(1)}$ is the effective Hamiltonian from Eq. \eqref{eq:Heff1ROAT}. In the following we will determine conditions under which this second-order term is small in order to justify our Taylor expansion.

\begin{figure}
    \centering
    \includegraphics[width=0.4\linewidth]{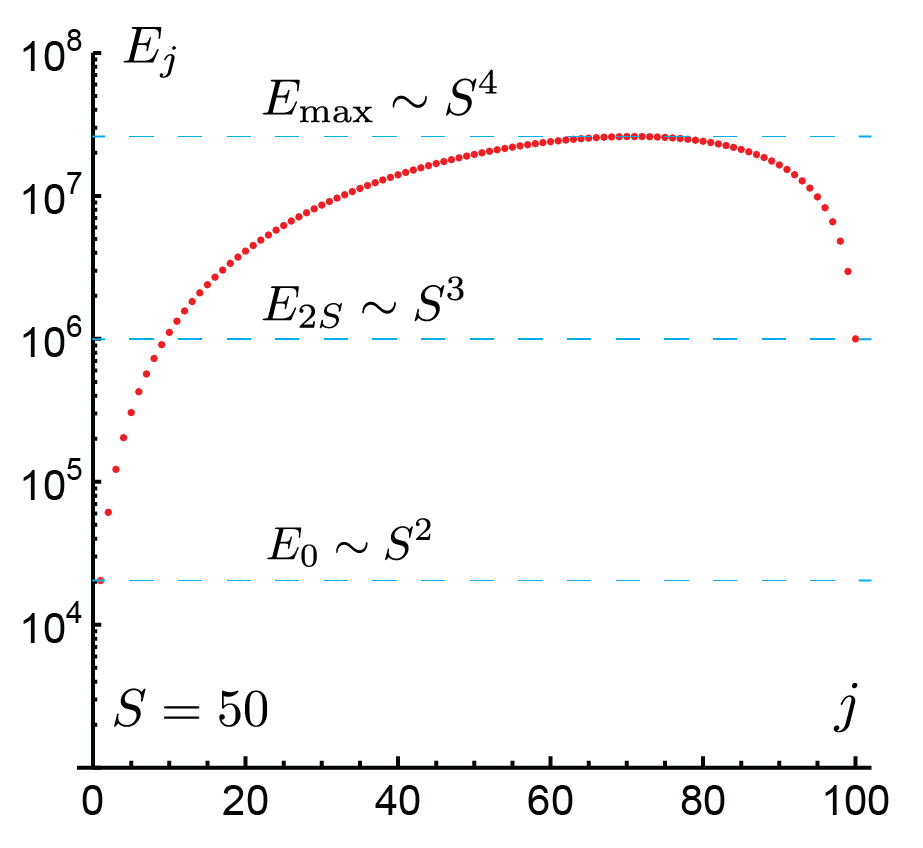}
    \caption{Spectrum of $H_{\mathrm{eff,R}}^{(1)}$ (red) for $S = 50$ as a function of $0 \leq j \leq 2S$ showing three energy scales (dotted blue).}
    \label{fig:k1spectrumscaling}
\end{figure}

The spectrum of $H_{\mathrm{eff,R}}^{(1)}$ is
\begin{equation}
    E_{\mathrm{eff,R}}^{(1)} = \chi \left(\frac{\chi \Delta t}{30} \right) j(j+1)(4S(S+1)-j(j+1)-1)
\end{equation}
where the total spin can take values $0 \leq j \leq 2S$. This spectrum features three energy scales depending on the value of $j$ as illustrated in Fig. \ref{fig:k1spectrumscaling}. For intermediate values of $j \sim S$ the energies scale like $E_{\mathrm{max}} \sim S^4$; for $j \rightarrow 2S$ the energies scale as $E_{2S} \sim S^3$; and for $j \rightarrow 1$ the gap above the ground state $j=0$ scales as $E_0 \sim S^2$. These different energy scales will determine under which conditions our Taylor expansion is appropriate.

First consider a twisting strength $\chi \Delta t = c / S^2$ per pulse where $c \ll 1$ is a constant independent of $S$. In this case the Taylor expansion is justified for all energy scales because $E_{\mathrm{eff,R}}^{(1)} \Delta t \lesssim (\chi \Delta t)^2 E_{\mathrm{max}} \sim (\chi \Delta t)^2 S^4 = c^2 \ll 1$. How many of these twists do we need in order to reach a metrologically useful state? Consider $K = u S^2$ twists giving a total twist strength $\chi T = K \chi \Delta t = K c / S^2 = u c$ where $u \gg 1$ is an $O(1)$ number that is independent of $S$. The time-to-design is controlled by the smallest energy scale $E_0 \sim \chi (\chi \Delta t) S^2 = c \chi$, so to reach a design we need $T E_0 = c \chi T = u c^2 \gg 1$. So for this choice of twisting strength per pulse, we require a total twisting strength that is an order-1 number. This is an exceptionally large twisting strength, corresponding to the same twisting required to generate a GHZ state.

Can we achieve a metrologically useful state using a smaller total twisting strength? Consider instead a twisting strength $\chi \Delta t = c / S^{b}$ per pulse where $c \ll 1$ and $b < 2$ is a tunable parameter. In this case the Taylor expansion is no longer justified for all energy scales because $E_{\mathrm{eff,R}}^{(1)} \Delta t \lesssim (\chi \Delta t)^2 S^4 = c^2 S^{4-2b}$, which is not small for $b < 2$. However, if we restrict ourselves to low-energy states on the scale of $E_{2S}$ and smaller, then we may safely Taylor-expand so long as $b \geq 3/2$ because $(\chi \Delta t)^2 E_{2S} \sim (\chi \Delta t)^2 S^3 = c^2 S^{3-2b} \ll 1$ when $b \geq 3/2$. Assuming that this we are justified in ignoring the highest-energy states $E_{\mathrm{max}}$ (as we shall show in a moment), then we may apply $K = u S$ twisting pulses, giving a total twisting strength $\chi T = K \chi \Delta t = K c / S^b = uc S^{1-b}$, which gives $\chi T \sim uc / \sqrt{S}$ when $b = 3/2$. This amount of twisting yields a design because we have $T E_0 \sim K (\chi \Delta t)^2 S^2 = u c^2 S^{3-2b} = u c^2 \gg 1$ when $b = 3/2$. Hence, if we are justified in ignoring the highest-energy states $E_{\mathrm{max}}$ then we achieve a design using a total twisting strength $\chi T \sim 1/\sqrt{S}$, which matches the total twisting strength found in the numerical analysis in the main text.

To eliminate the high energy states $\sim E_{\mathrm{max}} \sim S^4$ while still allowing for a Taylor expansion, we consider a two-phase scrambling protocol. In the first phase we apply $K_1 = u_1 S$ pulses with twisting strength $\chi \Delta t = c_1 / S^2$ per pulse. This choice allows for a Taylor expansion and eliminates all high-energy states lying above $E_{2S} \sim S^3$ because $T_1 E_{2S} \sim K_1 (\chi \Delta t)^2 S^3 \sim u_1 c_1^2  \gg 1$. This phase requires only a total twisting strength $\chi T_1 = K_1 \chi \Delta t = u_1 c_1 / S$, which is quite small. In the second phase we apply $K_2 = u_2 S$ twisting pulses with a twisting strength $\chi \Delta t = c_2 / S^{3/2}$ per pulse. At this point we are justified in using a Taylor expansion because the highest-energy states have been eliminated by the first phase, and we require only an additional total twisting strength $\chi T_2 = K_2 \chi \Delta t = u_2 c_2 /\sqrt{S}$. This total twisting strength matches the twisting strength found by our numerical results in the main text.

\section{Comparison to Minimum Mean Square Error Estimator}\label{app:mmse}
Here, we compare our butterfly echo protocol with minimum mean square error (MMSE) estimator constructed from a Bayesian perspective. Following 
Ref.~\cite{personick1971application}, define
\begin{align*}
\Gamma := \int d\theta\, p(\theta) \mathcal{M}_\theta[\rho] \\
\eta := \int d\theta\, \theta p(\theta) \mathcal{M}_\theta[\rho],
\end{align*}
where $p(\theta)$ is the prior distribution for the value of the parameter $\theta$. Here, we assume a uniform prior over $\theta\in[0,\theta_\mathrm{max}]$.
Then, the MMSE estimator for $\theta$ is the expectation value of an observable $A$ satisfying
\begin{equation}
\frac{1}{2}\{\Gamma, A\}= \eta.
\end{equation}
Assuming $\Gamma > 0$, the solution to this equation is a unique, Hermitian observable given by
\begin{equation}
A = 2\int_0^\infty d\alpha\, e^{-\Gamma \alpha}\eta e^{-\Gamma\alpha}.
\end{equation}
Assuming a uniform prior we can evaluate, for small $\theta_\mathrm{max}$:
\begin{align}
\Gamma &= \frac{1}{\theta_\mathrm{max}} \int_0^{\theta_\mathrm{max}} d\theta \left[\rho +\frac{\theta^2}{3}\left(\sum_j S_j\rho S_j-\frac{1}{2}\{\rho, S_j^2\}\right)+\mathcal{O}(\theta^3)\right]\nonumber\\
&= \rho +\frac{\theta_\mathrm{max}^2}{9}\left(\sum_j S_j\rho S_j-\frac{1}{2}\{\rho, S_j^2\}\right)+\mathcal{O}(\theta^3_\mathrm{max}),
\end{align}
and
\begin{align}
\eta &= \frac{1}{\theta_\mathrm{max}} \int_0^{\theta_\mathrm{max}} d\theta\, \left[\rho\theta +\frac{\theta^3}{3}\left(\sum_j S_j\rho S_j-\frac{1}{2}\{\rho, S_j^2\}\right)+\mathcal{O}(\theta^4)\right]\nonumber\\
&= \frac{\theta_\mathrm{max}}{2}\rho +\mathcal{O}(\theta_\mathrm{max}^3),
\end{align}
where we used \cref{eq:channel-new,eq:depolarization}.

To leading order in $\theta_\mathrm{max}$, the computation of $A$ is almost trivial. Using the fact that $\rho=\rho^2$ for a pure state, one finds, up to corrections $\mathcal{O}(\theta_\mathrm{max}^3)$,
\begin{equation}
A = \theta_\mathrm{max} \rho \int_0^\infty d\alpha\, e^{-2\alpha} = \frac{\theta_\mathrm{max} \rho}{2}.
\end{equation}

Thus, measuring $A$ for the encoded state $\mathcal{M}_\theta[\rho]$ we obtain the MMSE estimate
\begin{align}
\langle A\rangle &=\frac{\theta_{\mathrm{max}}}{2}\mathrm{Tr}\big(\rho\mathcal{M}_\theta[\rho]\big)\nonumber \\
&=\frac{\theta_{\mathrm{max}}}{2} \big(1+\mathcal{D}_\theta[\rho]+\mathcal{O}(\theta^3)\big).
\end{align}
Observe the crucial dependence of the uniform prior: even when the true value of the parameter $\theta=0$ and, thus, $\mathcal{D}_\theta[\rho]=0$ this estimator returns $\theta_\mathrm{max}/2$. Thus, as we expect for a Bayesian estimator taking into account prior information, this estimator is biased. 

Up to a normalization factor of $\theta_\mathrm{max}/2$ (which biases the MMSE observable), the MMSE observable to first order is to simply compute the overlap of the encoded state $\mathcal{M}_\theta[\rho]$ with the initial state. This is equivalent to the estimator used in the butterfly echo protocol.

\end{document}